\documentclass[a4paper,11pt]{article}

\usepackage{jheppub_mod} 
\usepackage[T1]{fontenc} 

\usepackage{hyperref}
\usepackage{graphicx}
\usepackage{amsmath,amssymb,slashed}
\usepackage{booktabs,tabulary}
\usepackage{dsfont}
\usepackage{color}
\usepackage{multirow}
\usepackage{subcaption}
\usepackage{placeins}

\usepackage{mathabx} 

\usepackage{braket} 
\usepackage{bbold} 
\usepackage{csquotes} 

\newcommand{\LN}[1]{\textsc{#1}} 
\newcommand{\Lat}[1]{\textit{#1}} 

\newcommand{\beq}{\begin{equation}}
\newcommand{\eeq}{\end{equation}}

\newcommand{\Ax}{A} 

\newcommand{\Maxial}{m_{\Ax}}
\newcommand{\Mf}{m_{f_1}}
\newcommand{\Mfprime}{m_{f_1'}}
\newcommand{\Ma}{m_{a_1}}

\newcommand{\MV}{M_{V}}
\newcommand{\MVprime}{M_{V'}}
\newcommand{\MVprimeprime}{M_{V''}}

\newcommand{\Mrho}{M_{\rho}}
\newcommand{\Mrhoprime}{M_{\rho'}}
\newcommand{\Mrhoprimeprime}{M_{\rho''}}
\newcommand{\Gammarho}{\Gamma_{\rho}}
\newcommand{\Gammarhoprime}{\Gamma_{\rho'}}
\newcommand{\Gammarhoprimeprime}{\Gamma_{\rho''}}

\newcommand{\Momega}{M_{\omega}}
\newcommand{\Momegaprime}{M_{\omega'}}
\newcommand{\Momegaprimeprime}{M_{\omega''}}

\newcommand{\Mphi}{M_{\phi}}
\newcommand{\Mphiprime}{M_{\phi'}}
\newcommand{\Mphiprimeprime}{M_{\phi''}}

\newcommand{\Mpi}{M_{\pi}}

\newcommand{\M}{\mathcal{M}}

\newcommand{\F}{\mathcal{F}}

\newcommand{\BR}{B}

\newcommand{\GeV}{\,\text{GeV}}
\newcommand{\MeV}{\,\text{MeV}}
\newcommand{\keV}{\,\text{keV}}

\newcommand{\perc}{\%}

\newcommand{\iu}{i}

\newcommand{\unity}{\mathds{1}}

\newcommand{\Order}{\mathcal{O}}

\newcommand{\dx}{\mathrm{d}x\,}
\newcommand{\dy}{\mathrm{d}y\,}
\newcommand{\dz}{\mathrm{d}z\,}

\renewcommand{\Re}{\text{Re}\,}

\newcommand{\Uthree}{U(3)}

\renewcommand{\vec}[1]{\boldsymbol{#1}} 

\newcommand{\eps}{\epsilon}

\newcommand{\aT}{\text{a}}
\newcommand{\sT}{\text{s}}

\newcommand{\asym}{\text{asym}}
\newcommand{\eff}{\text{eff}}

\newcommand{\sm}{s_\text{m}}

\allowdisplaybreaks[1]

\title{Axial-vector transition form factors and $\boldsymbol{e^+ e^- \to f_1 \pi^+ \pi^-}$}

\author[a]{Martin Hoferichter,}
\author[b]{Bastian Kubis,}
\author[b]{and Marvin Zanke}

\affiliation[a]{
Albert Einstein Center for Fundamental Physics, Institute for Theoretical Physics, University of Bern, Sidlerstrasse 5, 3012 Bern, Switzerland}

\affiliation[b]{
Helmholtz-Institut f\"ur Strahlen- und Kernphysik (Theorie) and \\
Bethe Center for Theoretical Physics, Universit\"at Bonn, 53115 Bonn, Germany}

\emailAdd{hoferichter@itp.unibe.ch}
\emailAdd{kubis@hiskp.uni-bonn.de}
\emailAdd{zanke@hiskp.uni-bonn.de}

\abstract{We study the transition form factors (TFFs) of axial-vector mesons in the context of currently available experimental data, including new constraints from $e^+ e^- \to f_1(1285) \pi^+ \pi^-$ that imply stringent limits on the high-energy behavior and, for the first time, allow us to provide an unambiguous determination of the couplings corresponding to the two antisymmetric TFFs. We discuss how these constraints can be implemented in a vector-meson-dominance picture, and, in combination with contributions from the light-cone expansion, construct TFFs as input for the evaluation of axial-vector contributions to hadronic light-by-light scattering in the anomalous magnetic moment of the muon.
}

\begin{document} 
\renewcommand{\figureautorefname}{Fig.}
\renewcommand{\tableautorefname}{Table}
\renewcommand{\chapterautorefname}{Ch.}
\renewcommand{\sectionautorefname}{Sec.}
\renewcommand{\subsectionautorefname}{Sec.}
\renewcommand{\appendixautorefname}{App.}
\def\equationautorefname~#1\null{Eq.~(#1)\null}
\maketitle

\section{Introduction}
\label{sec:intro}
The transition form factors (TFFs) of axial-vector mesons are crucial for estimating the hadronic light-by-light (HLbL) contribution to the anomalous magnetic moment of the muon $a_\mu$, in particular, for intermediate photon virtualities and the transition to short-distance constraints (SDCs).
At present, the axial-vector contribution included in the Standard-Model prediction for $a_\mu$~\cite{Aoyama:2020ynm,Aoyama:2012wk,Aoyama:2019ryr,Czarnecki:2002nt,Gnendiger:2013pva,Davier:2017zfy,Keshavarzi:2018mgv,Colangelo:2018mtw,Hoferichter:2019gzf,Davier:2019can,Keshavarzi:2019abf,Hoid:2020xjs,Kurz:2014wya,Melnikov:2003xd,Colangelo:2014dfa,Colangelo:2014pva,Colangelo:2015ama,Masjuan:2017tvw,Colangelo:2017qdm,Colangelo:2017fiz,Hoferichter:2018dmo,Hoferichter:2018kwz,Gerardin:2019vio,Bijnens:2019ghy,Colangelo:2019lpu,Colangelo:2019uex,Blum:2019ugy,Colangelo:2014qya} is responsible for a large fraction of the final uncertainty, $a_\mu^\text{HLbL} = 92(19) \times 10^{-11}$~\cite{Aoyama:2020ynm,Melnikov:2003xd,Masjuan:2017tvw,Colangelo:2017qdm,Colangelo:2017fiz,Hoferichter:2018dmo,Hoferichter:2018kwz,Gerardin:2019vio,Bijnens:2019ghy,Colangelo:2019lpu,Colangelo:2019uex,Pauk:2014rta,Danilkin:2016hnh,Jegerlehner:2017gek,Knecht:2018sci,Eichmann:2019bqf,Roig:2019reh}, especially, when taking into account the interplay with SDCs. 
In view of the expected experimental improvements beyond the current world average~\cite{Muong-2:2021ojo,Muong-2:2021ovs,Muong-2:2021xzz,Muong-2:2021vma,Muong-2:2006rrc}, the uncertainty in the HLbL contribution should be reduced by another factor of $2$ to ensure that it does not play a role in the interpretation of the experiment~\cite{Muong-2:2015xgu,Colangelo:2022jxc}. 
Such improvements are ongoing, both in lattice QCD~\cite{Chao:2021tvp,Chao:2022xzg,Blum:2023vlm,Alexandrou:2022qyf,Gerardin:2023naa} and with data-driven methods, including the derivation of higher-order SDCs~\cite{Bijnens:2020xnl,Bijnens:2021jqo,Bijnens:2022itw}, their implementation~\cite{Leutgeb:2019gbz,Cappiello:2019hwh,Knecht:2020xyr,Masjuan:2020jsf,Ludtke:2020moa,Colangelo:2021nkr,Leutgeb:2022lqw}, and dispersion relations~\cite{Hoferichter:2019nlq,Danilkin:2021icn,Holz:2022hwz,Ludtke:2023hvz}.
In particular, to evaluate the axial-vector contributions, one needs robust input for their TFFs. 
To this end, constraints on the asymptotic behavior have been derived from the light-cone expansion (LCE)~\cite{Hoferichter:2020lap} and available data evaluated in a vector-meson-dominance (VMD) inspired parameterization~\cite{Zanke:2021wiq}, but, due to scarcity of data, no sufficiently complete and reliable solutions have been obtained so far, with ambiguities mostly affecting the determination of the two antisymmetric TFFs. 

This is in large part because the interaction of an axial-vector resonance $\Ax$ with two electromagnetic currents is suppressed by the \LN{Landau}--\LN{Yang} theorem~\cite{Landau:1948kw,Yang:1950rg}---stating that a spin-$1$ particle cannot decay into two on-shell photons---so that all observables require at least one non-zero virtuality. 
Some data are available for the space-like process $e^+ e^- \to e^+ e^- \Ax$, for $\Ax = f_1 \equiv f_1(1285)$ and $\Ax = f_1' \equiv f_1(1420)$~\cite{Gidal:1987bn,Gidal:1987bm,TPCTwoGamma:1988izb,TPC-TWO-GAMMA:1988aph,Achard:2001uu,Achard:2007hm}, allowing one to extract the equivalent two-photon decay widths $\widetilde{\Gamma}_{\gamma\gamma}$ and, thereby, the mixing angle between the two $f_1$ states.
Accordingly, if $\Uthree$ symmetry is assumed, it suffices to determine the TFFs for the $f_1(1285)$ to be able to estimate the effect of the entire triplet including the $a_1(1260)$.
Focusing, therefore, on the $f_1$ resonance, for which most data are available, we compiled the constraints that follow from the radiative decays $f_1 \to \rho \gamma$ and $f_1 \to \phi \gamma$, as well as $f_1 \to 4\pi$ and $f_1 \to e^+ e^-$ in Ref.~\cite{Zanke:2021wiq}, improving on previous work~\cite{Rudenko:2017bel,Milstein:2019yvz} by employing parameterizations that ensure the absence of kinematic singularities, include SDCs, and incorporate the spectral functions of the isovector resonances. 
We found that, unfortunately, the decay $f_1 \to 4\pi$ does not provide any meaningful input for the TFFs, since dominated by $f_1 \to a_1 \pi \to \rho \pi \pi \to 4\pi$, while $f_1 \to e^+ e^-$ would, in principle, be a very interesting observable, yet not at the current level of precision~\cite{SND:2019rmq}. 
 
Ultimately, the currently available data were not sufficient to identify a unique solution for all three TFFs, especially the normalizations of the two antisymmetric TFFs and the momentum dependence of all three TFFs were only poorly determined.
In the future, these limitations could be overcome by better data for $e^+ e^- \to e^+ e^- f_1$ and $f_1 \to e^+ e^-$; in this paper, we instead propose to study existing data for $e^+ e^- \to f_1 \pi^+ \pi^-$~\cite{BaBar:2007qju,BaBar:2022ahi}.\footnote{%
    The measurement of the $e^+ e^- \to 2(\pi^+ \pi^-) \eta$~\cite{BaBar:2007qju} and $e^+ e^- \to K_S K^\pm \pi^\mp \pi^+ \pi^-$~\cite{BaBar:2022ahi} cross sections, in which the $f_1$ peak can be identified, is partly motivated by hadronic vacuum polarization (HVP).
    The impact of such high-multiplicity channels on the HVP contribution to $a_\mu$, though, is much smaller than the current tensions observed between data-driven evaluations~\cite{Davier:2017zfy,Keshavarzi:2018mgv,Colangelo:2018mtw,Hoferichter:2019gzf,Davier:2019can,Keshavarzi:2019abf,Hoid:2020xjs,Stamen:2022uqh,Colangelo:2022vok,Colangelo:2022prz,Hoferichter:2023sli,Hoferichter:2023bjm} and lattice QCD~\cite{Borsanyi:2020mff,Ce:2022kxy,ExtendedTwistedMass:2022jpw,FermilabLatticeHPQCD:2023jof,Blum:2023qou}, \Lat{e.g.}, $a_\mu^\text{HVP}[2(\pi^+ \pi^-) \eta] = 0.8(1) \times 10^{-11}$~\cite{Keshavarzi:2019abf} is at the same level as potential uncertainties from $\Order(\alpha^4)$ hadronic corrections~\cite{Hoferichter:2021wyj}.
    The CMD-3 measurement of $e^+ e^- \to 3(\pi^+ \pi^-) \pi^0$~\cite{CMD-3:2019ufp} includes results for $e^+ e^-\to 2(\pi^+ \pi^-) \eta$, but no additional information on $e^+ e^- \to f_1 \pi^+ \pi^-$.}
This process is also sensitive to all three TFFs, for one photon virtuality centered at the $\rho$ mass and the other one determined by the center-of-mass energy of the $e^+ e^-$ pair.
Phenomenologically, the reaction displays prominent resonance features from excited $\rho$ resonances~\cite{Liu:2022yrt}, primarily the $\rho(2150)$, but, when interpreted as a limit on the non-resonant contribution, entails powerful constraints on the TFFs of the $f_1$, both on the asymptotic behavior and the respective normalizations.
 
The outline of the paper is as follows: we first review the basic formalism for the axial-vector TFFs in \autoref{sec:TFFs}, and then define improved VMD parameterizations in \autoref{sec:VMD} that implement the asymptotic behavior observed in $e^+ e^- \to f_1 \pi^+ \pi^-$.
In \autoref{sec:observables}, we summarize the previous observables and present the formalism in which we will analyze $e^+ e^- \to f_1 \pi^+ \pi^-$.
The phenomenological analysis, including a review of the data base, a global fit, and a summary of the resulting TFF parameterizations, will be presented in \autoref{sec:pheno}, before concluding in \autoref{sec:summary}. 
Finally, in \autoref{appx:constants}, we collect constants and parameters used throughout this work.

\section{Axial-vector transition form factors}
\label{sec:TFFs}
The helicity amplitudes for the decay of an axial-vector meson into two virtual photons, $\Ax(P,\lambda_\Ax) \to \gamma^*(q_1,\lambda_1) \gamma^*(q_2,\lambda_2)$, are given by~\cite{Hoferichter:2020lap,Zanke:2021wiq}
\beq
    \M\big(\{\Ax,\lambda_\Ax\} \to \{\gamma^*,\lambda_1\} \{\gamma^*,\lambda_2\}\big) 
    = e^2 {\eps_\mu^{\lambda_1}}^*(q_1) {\eps_\nu^{\lambda_2}}^*(q_2) \eps_\alpha^{\lambda_\Ax}(P) \M^{\mu \nu \alpha}(q_1,q_2),
\eeq
where, following the \LN{Bardeen}--\LN{Tung}--\LN{Tarrach} procedure~\cite{Bardeen:1969aw,Tarrach:1975tu}, the tensor matrix element $\M^{\mu \nu \alpha}(q_1,q_2)$ can be decomposed into three independent \LN{Lorentz} structures and form factors $\F_i(q_1^2,q_2^2)$ that are free of kinematic singularities according to
\beq\label{eq:M_decomposition}
	\M^{\mu \nu \alpha}(q_1,q_2) 
    = \frac{\iu}{\Maxial^2} \sum_{i = \aT_1,\aT_2,\sT}{T_i^{\mu \nu \alpha}(q_1,q_2) \F_i(q_1^2,q_2^2)}.
\eeq
Here, $\Maxial$ is the mass of the respective axial-vector meson and the structures
\begin{align}
	T_{\aT_1}^{\mu \nu \alpha}(q_1,q_2)
	&= \eps^{\mu \nu \beta \gamma} {q_1}_\beta {q_2}_\gamma (q_1^\alpha - q_2^\alpha), \notag \\
	T_{\aT_2}^{\mu \nu \alpha}(q_1,q_2)
	&= \frac12 {q_1}_\beta {q_2}_\gamma \left(\eps^{\alpha \nu \beta \gamma} q_1^\mu + \eps^{\alpha \mu \beta \gamma} q_2^\nu\right) + \frac12 \eps^{\alpha \mu \nu \beta} ({q_2}_\beta q_1^2 + {q_1}_\beta q_2^2), \notag \\
	T_{\sT}^{\mu \nu \alpha}(q_1,q_2)
	&= \frac12 {q_1}_\beta {q_2}_\gamma \left(\eps^{\alpha \nu \beta \gamma} q_1^\mu - \eps^{\alpha \mu \beta \gamma} q_2^\nu\right) + \frac12 \eps^{\alpha \mu \nu \beta} ({q_2}_\beta q_1^2 - {q_1}_\beta q_2^2)
\end{align}
are completely antisymmetric ($\aT$) or symmetric ($\sT$) under photon crossing ($\mu \leftrightarrow \nu$ and $q_1 \leftrightarrow q_2$); similarly, the associated form factors obey the indicated symmetry properties under the exchange of momenta, $q_1^2 \leftrightarrow q_2^2$.
Furthermore, the prefactor $\iu/\Maxial^2$ in \autoref{eq:M_decomposition} is chosen to obtain dimensionless TFFs with real-valued normalization and the \LN{Levi}-\LN{Civita} tensor is used in the convention $\eps^{0123} = +1$.
For the formulation of SDCs and the analysis of the L3 data for $e^+ e^- \to e^+ e^- f_1$, see \autoref{sec:L3}, it is also useful to consider the basis defined by 
\begin{align}\label{eq:FFs_native_basis}
	\F_1(q_1^2,q_2^2) 
    &= \F_{\aT_1}(q_1^2,q_2^2), \notag \\
	\F_2(q_1^2,q_2^2) 
    &= \frac{1}{2} \big[ \F_{\aT_2}(q_1^2,q_2^2) + \F_{\sT}(q_1^2,q_2^2) \big], \notag \\
	\F_3(q_1^2,q_2^2) 
    &= \frac{1}{2} \big[ \F_{\aT_2}(q_1^2,q_2^2) - \F_{\sT}(q_1^2,q_2^2) \big].
\end{align}

Since the \LN{Landau}--\LN{Yang} theorem~\cite{Landau:1948kw,Yang:1950rg} forbids the decay into two on-shell photons, one commonly defines the so-called equivalent two-photon decay width as~\cite{TPCTwoGamma:1988izb}
\beq
 	\widetilde{\Gamma}_{\gamma\gamma} 
    = \lim_{q_1^2 \to 0} \frac{1}{2} \frac{\Maxial^2}{q_1^2} \Gamma(\Ax \to \gamma_\text{L}^* \gamma_\text{T}),
\eeq
with one longitudinal quasi-real photon $\gamma_\text{L}^*$, the spin-averaged---longitudinal--transversal ($\text{LT}$)---width 
\beq
	\Gamma(\Ax \to \gamma_\text{L}^* \gamma_\text{T}) 
    = \frac{1}{3} \sum_{\substack{\lambda_\Ax = \{0,\pm\} \\ \lambda_2 = \pm}} \int{\mathrm{d}\Gamma_{\Ax \to \gamma^* \gamma^*}^{0 \lambda_2 | \lambda_\Ax}} \Big\rvert_{q_2^2 = 0},
\eeq
and the differential decay width for fixed polarization
\beq
	\mathrm{d}\Gamma_{\Ax \to \gamma^* \gamma^*}^{\lambda_1 \lambda_2 | \lambda_\Ax} 
    = \frac{1}{32 \pi^2 \Maxial^2} \frac{\sqrt{\lambda(\Maxial^2, q_1^2, q_2^2)}}{2\Maxial} \lvert \M(\{\Ax,\lambda_\Ax\} \to \{\gamma^*,\lambda_1\} \{\gamma^*,\lambda_2\}) \rvert^2 \mathrm{d}\Omega,
\eeq
where $\Omega$ is the center-of-mass solid angle and $\lambda(a,b,c) = a^2 + b^2 + c^2 - 2ab - 2ac - 2bc$ is the \LN{K\"all\'en} function.
In terms of the above form factors, one finds
\beq\label{eq:two_photon_decay_width}
	\widetilde{\Gamma}_{\gamma\gamma} 
    = \frac{\pi \alpha^2}{48} \Maxial \lvert \F_\sT(0,0) \rvert^2,
\eeq
where $\alpha = e^2/(4\pi)$ is the fine-structure constant.

The asymptotic behavior of the axial-vector TFFs can be analyzed by means of the LCE, leading to~\cite{Hoferichter:2020lap}
\begin{align}
	\F_{\aT_1}(q_1^2,q_2^2) 
    &= \Order(1/Q^6), \notag \\
	\F_{\aT_2}(q_1^2,q_2^2) 
    &= \frac{F_{f_1}^\eff \Mf^3}{Q^4} f_{\aT_2}(w) + \Order(1/Q^6), 
    &
    f_{\aT_2}(w) 
    &= \frac{3}{4w^3}\left(6 + \frac{3 - w^2}{w} \log\frac{1 - w}{1 + w}\right), \notag \\
	\F_{\sT}(q_1^2,q_2^2) 
    &= \frac{F_{f_1}^\eff \Mf^3}{Q^4} f_{\sT}(w) + \Order(1/Q^6),
    &
    f_{\sT}(w)
    &= -\frac{3}{2w^3}\left(2w + \log\frac{1 - w}{1 + w}\right),
\end{align}
where 
\begin{align}
	Q^2 
    &= \frac{q_1^2 + q_2^2}{2} \in [0,\infty),
    &
    w
    &= \frac{q_1^2 - q_2^2}{q_1^2 + q_2^2} \in [-1,1]
\end{align}
denote the average photon virtuality and asymmetry parameter, respectively.
In the above, we furthermore introduced the effective decay constant
\beq
	F_\Ax^\eff 
    = 4 \sum_a{C_a F_\Ax^a},
\eeq
with the decay constants $F_\Ax^a$ defined via
\beq
	\braket{0 | \bar{q}(0) \gamma_\mu \gamma_5 \frac{\lambda^a}{2} q(0) | \Ax(P,\lambda_\Ax)}
    = F_\Ax^a \Maxial \eps_\mu.
\eeq
The \LN{Gell-Mann} matrices $\lambda_a$ and the conveniently normalized unit matrix $\lambda_0 = \sqrt{2/3} \, \unity$ determine the flavor decomposition, with the flavor weights $C_a$ in the effective decay constant given by $C_0 = 2/(3 \sqrt{6})$, $C_3 = 1/6$, and $C_8 = 1/(6 \sqrt{3})$.
In the symmetric doubly-virtual direction, the $\Order(1/q^4)$ limits become ($\lambda \approx 1$)~\cite{Zanke:2021wiq}
\begin{align}
	\F_{\aT_2}(q^2,\lambda q^2)
    &= -\frac{6 F_{f_1}^\eff \Mf^3}{q^4} k(\lambda) + \Order(1/q^6),
    \qquad
	\F_{\sT}(q^2,q^2)
    = \frac{F_{f_1}^\eff \Mf^3}{q^4} + \Order(1/q^6), \notag \\
	k(\lambda)
    &= \frac{3\lambda^2 - (\lambda^2 + 4 \lambda + 1) \log\lambda - 3}{(\lambda - 1)^4}
    = \Order(\lambda - 1),
\end{align}
but all singly-virtual limits of $\F_{\aT_2/\sT}(q_1^2,q_2^2)$ diverge in the symmetrized basis.
However, physical helicity amplitudes depend on linear combinations of the TFFs in such a way that only well-defined limits contribute to observables~\cite{Hoferichter:2020lap,Zanke:2021wiq}, which implies that $\F_2(q^2,0)$ and $\F_3(0,q^2)$ have finite limits, while the opposite cases diverge. 

\section{Vector-meson dominance}
\label{sec:VMD}
The parameterization of the TFFs inspired by VMD is based on their decomposition into isovector and isoscalar components.
Using $\Uthree$ symmetry, one can deduce the ratio~\cite{Zanke:2021wiq}
\beq
	R_{\text{S/V}}
    = \frac{\sqrt{2} - \tan\theta_\Ax}{3 (\sqrt{2} + \tan\theta_\Ax)}
    = -4.7(3.4)\perc
\eeq
of isoscalar to isovector contributions for the $f_1 \gamma^* \gamma^*$ coupling, where we inserted the L3 mixing angle $\theta_\Ax = 62(5)^\circ$~\cite{Achard:2001uu,Achard:2007hm},\footnote{%
    This determination of $\theta_\Ax$ assumes $\BR(f_1' \to K \bar K \pi) = 1$, supported by $\Gamma(f_1' \to \eta \pi \pi)/\Gamma(f_1' \to K \bar K \pi) < 0.1$~\cite{WA76:1991kit} and $\Gamma(f_1' \to a_0(980) \pi)/\Gamma(f_1' \to K \bar K \pi) = 0.040(14)$~\cite{WA102:1998zhh}.} 
derived from
\beq\label{eq:thetaA}
    \frac{\widetilde{\Gamma}_{\gamma\gamma}^{f_1}}{\widetilde{\Gamma}_{\gamma\gamma}^{f_1'}}
    = \frac{\Mf}{\Mfprime} \cot^2(\theta_\Ax - \theta_0),
    \qquad
    \theta_0
    = \arcsin\frac{1}{3},
\eeq
for the corresponding $J^{PC} = 1^{++}$ axial-vector nonet with the mixing pattern
\beq
	\begin{pmatrix} 
        f_1 \\
        f_1'
    \end{pmatrix} 
    =
    \begin{pmatrix} 
        \cos\theta_\Ax & \sin\theta_\Ax \\
        -\sin\theta_\Ax & \cos\theta_\Ax 
    \end{pmatrix} 
    \begin{pmatrix} 
        f^0 \\
        f^8 
    \end{pmatrix}.
\eeq
Hence, it is the isovector channel that dominates the process, with small isoscalar corrections at the level of $5\perc$.

The minimal particle content necessary for a VMD construction of TFFs that individually obey the asymptotic constraints summarized in \autoref{sec:TFFs} requires the inclusion of three multiplets.
More specifically, we will use $\rho \equiv \rho(770)$, $\rho' \equiv \rho(1450)$, and $\rho'' \equiv \rho(1700)$ for the isovector contributions and $\omega \equiv \omega(782)$, $\omega' \equiv \omega(1420)$, $\omega'' \equiv \omega(1650)$ as well as $\phi \equiv \phi(1020)$, $\phi' \equiv \phi(1680)$, $\phi'' \equiv \phi(2170)$ for the isoscalar contributions.
The introduction of a third multiplet, as required to obtain the correct asymptotic behavior for the antisymmetric TFFs, goes beyond the parameterizations of Ref.~\cite{Zanke:2021wiq}, ultimately, because the data on $e^+ e^- \to f_1 \pi^+ \pi^-$ demand such a steep decrease, including in kinematic configurations in which one virtuality is kept fixed at a finite but non-zero value.

\subsection{Isovector contributions}
In the space-like region, $q_i^2 < 0$, we propose to extend the isovector parameterizations from Ref.~\cite{Zanke:2021wiq} as follows: 
\begin{align}\label{eq:VMD_isovector}
	\F_{\aT_{1/2}}^{I=1}(q_1^2,q_2^2) 
    &= C_{\aT_{1/2}} \bigg[ \frac{(1 - \eps_{\aT_{1/2}}^{(1)} - \eps_{\aT_{1/2}}^{(2)}) \Mrho^2 \Mrhoprime^2}{(q_1^2 - \Mrho^2)(q_2^2 - \Mrhoprime^2)} + \frac{\eps_{\aT_{1/2}}^{(1)} \Mrho^2 \Mrhoprimeprime^2}{(q_1^2 - \Mrho^2)(q_2^2 - \Mrhoprimeprime^2)} \notag \\
    &\qquad\qquad + \frac{\eps_{\aT_{1/2}}^{(2)} \Mrhoprime^2 \Mrhoprimeprime^2}{(q_1^2 - \Mrhoprime^2)(q_2^2 - \Mrhoprimeprime^2)} \bigg] - (q_1 \leftrightarrow q_2), \notag \\
    \F_{\sT}^{I=1}(q_1^2,q_2^2) 
    &= C_{\sT} \bigg[ \frac{(1 - \eps_\sT^{(1)} - \eps_\sT^{(2)}) \Mrho^4}{(q_1^2 - \Mrho^2)(q_2^2 - \Mrho^2)} + \frac{(\eps_\sT^{(1)}/2) \Mrho^2 \Mrhoprime^2}{(q_1^2 - \Mrho^2)(q_2^2 - \Mrhoprime^2)} \notag \\
    &\qquad\quad + \frac{(\eps_\sT^{(1)}/2) \Mrhoprime^2 \Mrho^2}{(q_1^2 - \Mrhoprime^2)(q_2^2 - \Mrho^2)} + \frac{\eps_\sT^{(2)} \Mrhoprime^4}{(q_1^2 - \Mrhoprime^2)(q_2^2 - \Mrhoprime^2)} \bigg],
\end{align}
first given in this form to emphasize that, upon a partial-fraction decomposition, each term corresponds to adding vector-meson propagators with fixed coefficients.
To implement the correct singly-virtual asymptotic behavior, we choose
\begin{align}
    \eps_{\aT_{1/2}}^{(1)}
    &= -\frac{\Mrhoprime^2}{\Mrhoprimeprime^2 - \Mrhoprime^2 + \Mrho^2},
    &
    \eps_{\aT_{1/2}}^{(2)}
    &= \frac{\Mrho^2}{\Mrhoprimeprime^2 - \Mrhoprime^2 + \Mrho^2}, \notag \\
    \eps_\sT^{(1)}
    &= -\frac{2 \Mrho^2 \Mrhoprime^2}{(\Mrhoprime^2 - \Mrho^2)^2},
    &
    \eps_\sT^{(2)}
    &= \frac{\Mrho^4}{(\Mrhoprime^2 - \Mrho^2)^2},
\end{align}
leading to
\begin{align}\label{eq:VMD_isovector_inserted}
    \F_{\aT_{1/2}}^{I=1}(q_1^2,q_2^2)
    &= \frac{C_{\aT_{1/2}} \zeta_\rho \Mrho^4 \Mrhoprime^4 \Mrhoprimeprime^4 (q_1^2 - q_2^2)}{(q_1^2 - \Mrho^2)(q_2^2 - \Mrho^2)(q_1^2 - \Mrhoprime^2)(q_2^2 - \Mrhoprime^2)(q_1^2 - \Mrhoprimeprime^2)(q_2^2 - \Mrhoprimeprime^2)}, \notag \\
    \F_{\sT}^{I=1}(q_1^2,q_2^2)
    &= \frac{C_{\sT} \Mrho^4 \Mrhoprime^4}{(q_1^2 - \Mrho^2)(q_2^2 - \Mrho^2)(q_1^2 - \Mrhoprime^2)(q_2^2 - \Mrhoprime^2)},
\end{align}
with
\beq\label{eq:zetaV}
    \zeta_V
    = \frac{(\MVprimeprime^2 - \MVprime^2)(\MVprimeprime^2 - \MV^2)(\MVprime^2 - \MV^2)}{\MVprimeprime^2 \MVprime^2 \MV^2 (\MVprimeprime^2 - \MVprime^2 + \MV^2)}.
\eeq
The resulting asymptotic behavior of the TFFs becomes
\begin{align}\label{eq:VMD_asymptotics}
    \F_{\aT_{1/2}}^{I=1}(q_1^2,q_2^2) 
    & \ \propto \ \frac{1}{q_2^4},
    &
    \F_{\aT_{1/2}}^{I=1}(q^2,\lambda q^2) 
    & \ \propto \ \frac{1 - \lambda}{\lambda^3} \frac{1}{q^{10}}, \notag \\
    \F_\sT^{I=1}(q_1^2,q_2^2) 
    & \ \propto \ \frac{1}{q_2^4},
    &
    \F_\sT^{I=1}(q^2,\lambda q^2) 
    & \ \propto \ \frac{1}{q^8},
\end{align}
with $q_1^2$ fixed to a finite value distinct from $q_2^2$ (left) and in the doubly-virtual direction (right).
Crucially, the singly-virtual asymptotics now match the LCE result from \autoref{sec:TFFs} for arbitrary fixed $q_1^2$, which, for $q^2_1 = \Mrho^2$, is mandatory for a realistic description of the $e^+ e^- \to f_1\pi^+ \pi^-$ data (the opposite case with fixed $q_2^2$ follows from symmetry).\footnote{%
    For $\F_{\aT_{1}}^{I=1}(q_1^2,q_2^2)$, the LCE predicts an even faster decrease in the singly-virtual direction, but we do not consider yet another multiplet for the following reasons: (i) information from the LCE on this TFF is limited, \Lat{i.e.}, no non-vanishing contribution survives at $\Order(1/Q^4)$, in such a way that, in contrast to the other TFFs, we cannot add an LCE term to repair the behavior in the doubly-virtual direction and thus need to choose a compromise; (ii) the fit to $e^+ e^- \to f_1 \pi^+ \pi^-$ produces a small coupling $C_{\aT_{1}}$, in line with the LCE suppression; (iii) another multiplet would have a mass already in the energy range in which the data are to be described, so that no meaningful suppression could be generated even when introducing another state.}
For time-like applications, the replacements $M^2 \to M^2 - \iu M \Gamma$ apply in the denominators of \autoref{eq:VMD_isovector_inserted}, \Lat{i.e.}, after imposing the asymptotic behavior of the TFFs; due to the large widths of the $\rho$-like mesons, a narrow-width approximation, $M^2 \to M^2 - \iu \eps$ in the denominators, in general becomes insufficient here.
A consequence of the faster decrease in the singly-virtual directions concerns an even faster decrease in the doubly-virtual case, much below the LCE expectation. 
Accordingly, in the final representation for the TFFs, we add the asymptotic contribution~\cite{Zanke:2021wiq}
\begin{align}
	\F_{\aT_2}^\asym(q_1^2,q_2^2) 
	&= 3 F_{f_1}^\eff \Mf^3\big(q_1^2 - q_2^2\big) \int_{\sm}^\infty{\dx \frac{q_1^2 q_2^2 - x^2 + x (q_1^2 + q_2^2)}{(x - q_1^2)^3 (x - q_2^2)^3}}, \notag \\
	\F_\sT^\asym(q_1^2,q_2^2)
    &= 3 F_{f_1}^\eff \Mf^3 \int_{\sm}^\infty{\dx \frac{(q_1^2 + q_2^2)(x^2 - q_1^2 q_2^2) - x(q_1^2 - q_2^2)^2}{(x - q_1^2)^3 (x - q_2^2)^3}},
\end{align}
where $\sm$ is a parameter that determines the scale of the transition.
The implementation of these asymptotic contributions, or their variant including mass effects~\cite{Zanke:2021wiq}, becomes relevant for the axial-vector contributions in the HLbL loop integral.
Here, we focus on the determination of the low-energy couplings in the VMD component of the parameterization, as can be obtained from $e^+ e^- \to f_1\pi^+ \pi^-$.

\subsection{Isoscalar contributions}
In complete analogy to the above, the isoscalar parts of the form factors are parameterized according to\footnote{%
    We assume ideal mixing for the vector mesons, which prevents crossed terms involving $\omega$ and $\phi$ states.}
\begin{align}\label{eq:VMD_isoscalar}
    \F_{\aT_{1/2}}^{I=0}(q_1^2,q_2^2)
    &= \sum_{V = \omega,\phi} \frac{C_{\aT_{1/2}}^V \zeta_V \MV^4 \MVprime^4 \MVprimeprime^4 (q_1^2 - q_2^2)}{(q_1^2 - \MV^2)(q_2^2 - \MV^2)(q_1^2 - \MVprime^2)(q_2^2 - \MVprime^2)(q_1^2 - \MVprimeprime^2)(q_2^2 - \MVprimeprime^2)}, \notag \\
    \F_{\sT}^{I=0}(q_1^2,q_2^2)
    &= \sum_{V = \omega,\phi} \frac{C_{\sT}^V \MV^4 \MVprime^4}{(q_1^2 - \MV^2)(q_2^2 - \MV^2)(q_1^2 - \MVprime^2)(q_2^2 - \MVprime^2)},
\end{align}
with the same asymptotic properties as in \autoref{eq:VMD_asymptotics}.
Again, time-like applications imply the replacements $M^2 \to M^2 - \iu M \Gamma$ in the denominators, since the large widths of the excited isoscalar resonances do not allow for a narrow-width approximation.
Finally, under the assumption of $\Uthree$ symmetry, the isoscalar coupling constants can be related to the isovector analogs, leading to the approximations~\cite{Zanke:2021wiq}
\begin{align}\label{eq:SU3_ratios_couplings}
    R^\omega
    &= \frac{C_{\aT_{1/2}}^\omega}{C_{\aT_{1/2}}} 
    = \frac{C_\sT^\omega}{C_\sT}
    = \frac{1}{9}, \notag \\
    R^\phi
    &= \frac{C_{\aT_{1/2}}^\phi}{C_{\aT_{1/2}}} 
    = \frac{C_\sT^\phi}{C_\sT}
    = \frac{2\sqrt{2}}{9} \cot(\theta_\Ax + \theta_1) 
    = -0.158(34),
\end{align}
with $\theta_1 = \arctan\sqrt{2} = (\pi + 2\theta_0)/4$.

\section{Observables}
\label{sec:observables}

\subsection[$e^+ e^- \to e^+ e^- f_1$]{$\boldsymbol{e^+ e^- \to e^+ e^- f_1}$}
\label{sec:L3}
The equivalent two-photon decay width $\widetilde{\Gamma}_{\gamma \gamma}^{f_1} = 3.5(6)(5) \keV$, as measured by the L3 collaboration~\cite{Achard:2001uu}, determines the normalization of the symmetric TFF, see \autoref{eq:two_photon_decay_width}.
Taking into account the isoscalar contributions, $\lvert \F_{\sT}^{I=1}(0,0) + \F_{\sT}^{I=0}(0,0)\rvert = (1 + R^\omega + R^\phi) \lvert C_{\sT}\rvert = 0.953(34) \lvert C_{\sT}\rvert$, we have
\beq\label{eq:Cs_isoscalar}
	C_{\sT}
    = 0.93(11),
\eeq
where we followed the sign convention of Ref.~\cite{Zanke:2021wiq}.

The singly-virtual VMD limits can be further constrained by matching the L3 parameterization onto the full description of the $e^+ e^- \to e^+ e^- f_1$ cross section~\cite{Hoferichter:2020lap},
\beq
    \bigg\lvert \bigg( 1 - \frac{q^2}{\Mf^2} \bigg) \F_1(q^2,0) - \frac{q^2}{\Mf^2} \F_2(q^2,0) \bigg\rvert^2 - \frac{2q^2}{\Mf^2} \big\lvert \F_2(q^2,0) \big\rvert^2
    = \frac{-q^2}{\Mf^2}\bigg(2 - \frac{q^2}{\Mf^2}\bigg) \lvert \F_\text{D}(q^2,0) \rvert^2,
\eeq
where
\beq
	\F_\text{D}(q^2,0)
    = \frac{\F_\text{D}(0,0)}{(1 - q^2/\Lambda_\text{D}^2)^2}
\eeq
is the dipole ansatz assumed in Ref.~\cite{Achard:2001uu} and the form factors are given in the basis of \autoref{eq:FFs_native_basis}. 
While the normalization agrees by construction, matching the slopes at $q^2 = 0$ leads to
\begin{align}
    \frac{2}{\Lambda_\text{D}^2}
    &= \frac{1}{N_{\omega\phi}} \Bigg[ \frac{1}{\Mrho^2} + \frac{1}{\Mrhoprime^2} + R^\omega \bigg( \frac{1}{\Momega^2} + \frac{1}{\Momegaprime^2} \bigg) + R^\phi \bigg( \frac{1}{\Mphi^2} + \frac{1}{\Mphiprime^2} \bigg) \\
    &\qquad\qquad + (\zeta_\rho + \zeta_\omega R^\omega + \zeta_\phi R^\phi) \frac{C_{\aT_1}+C_{\aT_2}}{C_\sT} - \frac{\Mf^2(\zeta_\rho + \zeta_\omega R^\omega + \zeta_\phi R^\phi)^2}{N_{\omega\phi}} \bigg( \frac{C_{\aT_1}}{C_{\sT}} \bigg)^2 \Bigg], \notag
\end{align}
where the factor $N_{\omega\phi} = 1 + R^\omega + R^\phi$ accounts for the isoscalar terms in the normalization. 

\subsection[$f_1 \to \rho \gamma$ and $f_1 \to \phi \gamma$]{$\boldsymbol{f_1 \to \rho \gamma}$ and $\boldsymbol{f_1 \to \phi \gamma}$}
From the procedure outlined in Ref.~\cite{Zanke:2021wiq}, it is straightforward to obtain the branching ratio of $f_1 \to V \gamma$, $V=\rho, \omega, \phi$, in the form
\beq
    \BR(f_1 \to V \gamma) 
    = (R^V)^2 \frac{B^V_1 \big( \widetilde{C}_{\aT_1}^V \big)^2 + B^V_2 \big( \widetilde{C}_{\aT_2}^V + \widetilde{C}_{\sT}^V \big)^2 - B^V_3 \widetilde{C}_{\aT_1}^V\big( \widetilde{C}_{\aT_2}^V + \widetilde{C}_{\sT}^V \big)}{\Gamma_f},
\eeq
where we defined
\begin{align}
	B^V_1 
    &= \frac{\alpha \lvert g_{V \gamma} \rvert^2 \big(\Mf^2 - M_V^2\big)^5}{24 \Mf^9}, 
    &
    B^V_2
    &= \frac{\alpha \lvert g_{V \gamma} \rvert^2 M_V^2 \big(\Mf^2 - M_V^2\big)^3 \big(\Mf^2 + M_V^2\big)}{96 \Mf^9}, \notag \\
	B^V_3 
    &= \frac{\alpha \lvert g_{V \gamma} \rvert^2 M_V^2 \big(\Mf^2 - M_V^2\big)^4}{24 \Mf^9}, 
\end{align}
and the couplings
\begin{align}
    \widetilde{C}_{\aT_{1/2}}^V 
    &= J_{\aT}^V C_{\aT_{1/2}},
    &
    \widetilde{C}_{\sT}^V
    &= J_{\sT}^V C_\sT
\end{align}
are rescaled by
\begin{align}
    J_{\aT}^V
    &= \frac{\MVprimeprime^2 - \MVprime^2}{\MVprimeprime^2 - \MVprime^2 + \MV^2}, 
    &
    J_{\sT}^V 
    &= \frac{\MVprime^2}{\MVprime^2 - \MV^2}.
\end{align}
The normalizations $R^V$, $V = \omega,\phi$, are given by \autoref{eq:SU3_ratios_couplings}, and $R^\rho = 1$.

In addition, for $V = \rho$, information on the helicity amplitudes is available.
The spin-averaged amplitude squared of the corresponding process $f_1 \to \rho \gamma \to \pi^+ \pi^- \gamma$, with the subsequent decay of an on-shell $\rho$ meson, is of the generic form
\beq
	\big\lvert \M(f_1 \to \rho \gamma \to \pi^+ \pi^- \gamma) \big\rvert^2
    = \mathrm{M}_{\text{TT}} \sin^2{\theta_{\pi^+\gamma}} + \mathrm{M}_{\text{LL}} \cos^2{\theta_{\pi^+ \gamma}},
\eeq
where $\theta_{\pi^+ \gamma}$ is the angle between the final-state $\pi^+$ and the photon.
Similarly to the branching ratio of $f_1 \to \rho \gamma$, the ratio of helicity amplitudes for $f_1 \to \rho \gamma \to \pi^+ \pi^- \gamma$ results from a straightforward modification of the result presented in Ref.~\cite{Zanke:2021wiq},
\beq
	r_{\rho \gamma} 
    = \frac{\mathrm{M}_{\text{LL}}}{\mathrm{M}_{\text{TT}}} 
    = \frac{2 \Mf^2 \Mrho^2}{\big[\Mrho^2 - 2 \big(\Mf^2 - \Mrho^2\big) \widetilde{C}_{\aT_1}/\big(\widetilde{C}_{\aT_2} + \widetilde{C}_{\sT}\big)\big]^2}.
\eeq

\subsection[$f_1 \to e^+ e^-$]{$\boldsymbol{f_1 \to e^+ e^-}$}
\begin{table}[t]
	\centering
	\begin{tabular}{l  c  c}
	\toprule
	& Narrow-resonance limit & Spectral function for $\rho$ \\\midrule
	$D_1^{I=1} \times 10^3$ & $0.23 - 1.68\iu$ & $0.15 - 1.42\iu$ \\
	$D_2^{I=1} \times 10^3$ & $-1.70 + 1.81\iu$ & $-1.36 + 1.64\iu$ \\
	$D_3^{I=1} \times 10^3$ & $1.25 + 4.38\iu$ & $3.00 + 4.10\iu$ \\
	$D_1^{\omega} \times 10^3$ & \multicolumn{2}{c}{$0.30 - 1.65\iu$} \\
	$D_2^{\omega} \times 10^3$ & \multicolumn{2}{c}{$-1.89 + 1.81\iu$} \\
	$D_3^{\omega} \times 10^3$ & \multicolumn{2}{c}{$1.06 + 4.61\iu$} \\
	$D_1^{\phi} \times 10^3$ & \multicolumn{2}{c}{$-0.98 - 1.09\iu$} \\
	$D_2^{\phi} \times 10^3$ & \multicolumn{2}{c}{$0.19 + 2.42\iu$} \\
	$D_3^{\phi} \times 10^3$ & \multicolumn{2}{c}{$6.02 + 5.97\iu$} \\
	$D_\asym \times 10^3$ & \multicolumn{2}{c}{$0.125(12)\quad 0.032(3)\quad 0.017(2)\quad 0.009(1)$} \\
	\bottomrule
	\end{tabular}
	\caption{Numerical values for the coefficients $D_i^I$ in \autoref{eq:A1_decomposition} (obtained using the \Lat{Cuba} library~\cite{Hahn:2004fe}).
    The total isoscalar one follows as $D_i^{I=0} = R^\omega D_i^\omega + R^\phi D_i^\phi$, and $D_\asym$ is given for the matching points $\sqrt{\sm} \in \{1.0,1.3,1.5,1.7\}\GeV$.
    The left column gives the reference point for which the widths of all vector mesons are neglected and the right column the more realistic case that includes the spectral function of the $\rho$ (used as input in \autoref{tab:solutions_global_fit}).}
	\label{tab:coefficients_Di}
\end{table}
We follow Ref.~\cite{Zanke:2021wiq} and write the decay rate for $f_1 \to e^+ e^-$ as
\beq
	\Gamma(f_1 \to e^+ e^-)
    = \frac{64 \pi^3 \alpha^4 \Mf}{3} \lvert A_1 \rvert^2,
\eeq
where the scalar amplitude $A_1$ is implicitly defined by
\beq
	\M(f_1(P) \to e^+(p_2) e^-(p_1)) 
    = e^4 \eps_\mu(P) \overline{u}^s(p_1) \gamma^\mu \gamma^5 A_1 v^r(p_2)
\eeq
and further decomposes into terms proportional to the three VMD couplings (with isoscalar and isovector coefficients $D_i^I$) and an asymptotic contribution $D_\asym$,
\beq\label{eq:A1_decomposition}
    A_1 
    = \big( D_1^{I=1} + D_1^{I=0} \big) C_{\aT_1} + \big( D_2^{I=1} + D_2^{I=0} \big) C_{\aT_2} + \big( D_3^{I=1} + D_3^{I=0} \big) C_{\sT} + D_\asym.
\eeq
For the case of products of narrow-resonance propagators as in \autoref{eq:VMD_isovector}, with squared masses $x$ and $y$, the integral representation
\beq
    D_i
    = \frac{x y}{16\pi^2 \Mf^4} \int_0^1 \dz f_i(x,y,z,\Mf)
\eeq
applies,
with analytic expressions for the functions $f_i(x,y,z,\Mf)$ given in Ref.~\cite{Zanke:2021wiq}.
In the same reference, we also provided evaluations of the asymptotic contribution $D_\asym$ and studied in detail the sensitivity of the integrals to the spectral functions assumed for $\rho$ and $\rho'$ resonances when going beyond a narrow-resonance picture, to avoid unphysical imaginary parts in the loop integrals~\cite{Lomon:2012pn,Moussallam:2013una,Crivellin:2022gfu}.
In the decay region of the $f_1$, by far the most important correction arises from the width of the $\rho$. 
Here, we provide a simple evaluation of $\BR(f_1 \to e^+ e^-)$ for the representations constructed in \autoref{sec:VMD} that captures this main effect, to ensure that our final solutions do not conflict with the SND measurement~\cite{SND:2019rmq}.
To this end, we replace 
\begin{align}
    \F_{\aT_{1/2}}^{I=1}(q_1^2,q_2^2)
    &\to \frac{C_{\aT_{1/2}}}{N_{\aT_{1/2}}} \frac{\Mrhoprime^2 \Mrhoprimeprime^2 (\Mrhoprimeprime^2 - \Mrhoprime^2)(q_1^2 - q_2^2)}{(q_1^2 - \Mrhoprime^2)(q_2^2 - \Mrhoprime^2)(q_1^2 - \Mrhoprimeprime^2)(q_2^2 - \Mrhoprimeprime^2)} \notag \\
    &\times \frac{1}{\pi} \int_{4\Mpi^2}^\infty \dx \frac{x (\Mrhoprimeprime^2 - x)(\Mrhoprime^2 - x)\rho(x)}{(\Mrhoprimeprime^2 - \Mrhoprime^2 + x)(q_1^2 - x)(q_2^2 - x)}, \notag \\
    \F_{\sT}^{I=1}(q_1^2,q_2^2)
    &\to \frac{C_{\sT}}{N_{\sT}^2} \frac{\Mrhoprime^4}{(q_1^2 - \Mrhoprime^2)(q_2^2 - \Mrhoprime^2)}\frac{1}{\pi^2} \int_{4\Mpi^2}^\infty \dx \int_{4\Mpi^2}^\infty \dy \frac{x y \rho(x)\rho(y)}{(q_1^2 - x)(q_2^2 - y)}
\end{align}
in \autoref{eq:VMD_isovector_inserted}, where the normalizations $N_{\aT_{1/2}}$, $N_{\sT}$ of the spectral function $\rho(x)$ (taken from Refs.~\cite{Zanke:2021wiq,COMPASS:2015gxz,VonHippel:1972fg}) are determined by demanding that the meaning of the couplings $C_{\aT_{1/2}}$, $C_{\sT}$ remain unaltered compared to the zero-width limit; numerical results for the coefficients $D_i$ are collected in \autoref{tab:coefficients_Di}.\footnote{%
    Besides the analytic evaluation using the functions $f_i(x,y,z,\Mf)$, we performed cross checks by means of a \LN{Passarino}--\LN{Veltman} decomposition, obtained with \Lat{FeynCalc}~\cite{Mertig:1990an,Shtabovenko:2016sxi,Shtabovenko:2020gxv}, and the subsequent calculation of the loop integrals with \Lat{Collier}~\cite{Denner:2002ii,Denner:2005nn,Denner:2010tr,Denner:2016kdg,Schafer:2023qtl}.} 
Once improved data on $\BR(f_1 \to e^+ e^-)$ become available, more refined analyses can be performed along the lines of Refs.~\cite{Zanke:2021wiq,Hoferichter:2021lct}.

\subsection[$e^+ e^- \to f_1 \rho$]{$\boldsymbol{e^+ e^- \to f_1 \rho}$}
\begin{figure}[t]
	\centering
	\includegraphics{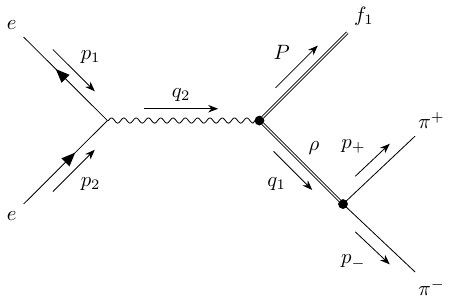}
	\caption{\LN{Feynman} diagram for $e^+ e^- \to f_1 \rho \to f_1 \pi^+ \pi^-$ consistent with $\M(\gamma^* \to f_1 \rho^{0^*})$.}
	\label{fig:ee_f1_pipi}
\end{figure}
The scattering process $e^+ e^- \to f_1 \pi^+ \pi^-$ probes the $f_1$ TFFs in the time-like region via $e^+ e^- \to \gamma^* \to f_1 \rho \to f_1 \pi^+ \pi^-$, see \autoref{fig:ee_f1_pipi}.
For our analysis, we thus define amputated $f_1 \to \rho \gamma^*$ form factors, which are related to $\gamma^* \to f_1 \rho$ via crossing symmetry, according to
\begin{align}
	\widebar{\F}_{\aT_{1/2}}^{I=1}(q_2^2) 
    &= C_{\aT_{1/2}} \bigg[ \frac{(1 - \eps_{\aT_{1/2}}^{(1)} - \eps_{\aT_{1/2}}^{(2)}) \Mrho^2 \Mrhoprime^2}{q_2^2 - \Mrhoprime^2} + \frac{\eps_{\aT_{1/2}}^{(1)} \Mrho^2 \Mrhoprimeprime^2}{q_2^2 - \Mrhoprimeprime^2} \bigg] 
    = -\frac{C_{\aT_{1/2}} \widebar{\zeta}_\aT}{(q_2^2 - \Mrhoprime^2)(q_2^2 - \Mrhoprimeprime^2)} \notag \\
    &\to -\frac{C_{\aT_{1/2}} \widebar{\zeta}_\aT}{(q_2^2 - \Mrhoprime^2 + \iu \Mrhoprime \Gammarhoprime)(q_2^2 - \Mrhoprimeprime^2 + \iu \Mrhoprimeprime \Gammarhoprimeprime)}, \notag \\
    \widebar{\F}_{\sT}^{I=1}(q_2^2) 
    &= C_{\sT} \bigg[ \frac{(1 - \eps_\sT^{(1)} - \eps_\sT^{(2)}) \Mrho^4}{q_2^2 - \Mrho^2} + \frac{(\eps_\sT^{(1)}/2) \Mrho^2 \Mrhoprime^2}{q_2^2 - \Mrhoprime^2} \bigg] 
    = -\frac{C_\sT \widebar{\zeta}_\sT}{(q_2^2 - \Mrho^2)(q_2^2 - \Mrhoprime^2)} \notag \\
    &\to -\frac{C_\sT \widebar{\zeta}_\sT}{(q_2^2 - \Mrho^2 + \iu \Mrho \Gammarho)(q_2^2 - \Mrhoprime^2 + \iu \Mrhoprime \Gammarhoprime)},
\end{align}
where 
\begin{align}
    \widebar{\zeta}_\aT
    &= \frac{\Mrho^2 \Mrhoprime^2 \Mrhoprimeprime^2 (\Mrhoprimeprime^2 - \Mrhoprime^2)}{\Mrhoprimeprime^2 - \Mrhoprime^2 + \Mrho^2},
    &
    \widebar{\zeta}_\sT
    &= \frac{\Mrho^4 \Mrhoprime^4}{\Mrhoprime^2 - \Mrho^2},
\end{align}
and a width has been inserted into the denominators for the time-like application.

\begin{figure}[t]
	\centering
	\includegraphics{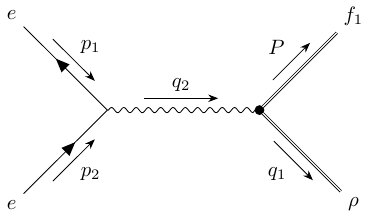}
	\caption{\LN{Feynman} diagram for $e^+ e^- \to f_1 \rho$ consistent with $\M(\gamma^* \to f_1 \rho^{0^*})$.}
	\label{fig:ee_f1_rho}
\end{figure}

As a first approximation, we consider the case in which the decay $e^+ e^- \to f_1 \pi^+ \pi^-$ is described by $e^+ e^- \to f_1 \rho$, see \autoref{fig:ee_f1_rho}, whose amplitude can be constructed from \autoref{eq:M_decomposition} and \autoref{eq:VMD_isovector} by amputating the $\rho$ propagator,
\begin{align}
    \M(\gamma^* \to f_1 \rho^{0^*})
    &= \frac{e}{\widetilde{g}_{\rho \gamma}\Mf^2} \eps_\mu^*(q_1) \eps_\nu(q_2) \eps_\alpha^*(P) \\
    &\hspace{-1cm} \times \Big[ T_{\aT_1}^{\mu \nu \alpha}(-q_1,q_2) \widebar{\F}_{\aT_1}(q_2^2) + T_{\aT_2}^{\mu \nu \alpha}(-q_1,q_2) \widebar{\F}_{\aT_2}(q_2^2) + T_{\sT}^{\mu \nu \alpha}(-q_1,q_2) \widebar{\F}_{\sT}(q_2^2) \Big], \notag
\end{align}
and the prefactor follows in analogy to the derivation of $f_1 \to V \gamma$ in Ref.~\cite{Zanke:2021wiq}.
In particular, we have taken over the definition
\beq
    \widetilde{g}_{V \gamma}
    = \frac{M_V^2}{g_{V \gamma}},
\eeq
to establish the connection to $g_{V \gamma}$, which, in a narrow-width approximation, is related to the dilepton decay
\beq\label{eq:G_V_gamma}
    \Gamma(V \to e^+ e^-)
    = \frac{4\pi \alpha^2}{3 \lvert g_{V \gamma} \rvert^2} \left(1 + \frac{2m_e^2}{M_V^2}\right) \sqrt{M_V^2 - 4m_e^2}\,,
\eeq
or, more rigorously, to the residue at the pole~\cite{Hoferichter:2017ftn}.
In order to determine the amplitude for $e^+ e^- \to \gamma^* \to f_1 \rho$, we calculate the diagram shown in \autoref{fig:ee_f1_rho}, leading to 
\begin{align}
    \M(e^+ e^- \to f_1 \rho)
    &= \frac{e^2}{\widetilde{g}_{\rho \gamma} \Mf^2} \eps_\mu^*(q_1) \eps_\alpha^*(P) \frac{\overline{v}^s(p_1) \gamma_\nu u^r(p_2)}{q_2^2} \\
    &\hspace{-1.75cm} \times \Big[ T_{\aT_1}^{\mu \nu \alpha}(-q_1,q_2) \widebar{\F}_{\aT_1}(q_2^2) + T_{\aT_2}^{\mu \nu \alpha}(-q_1,q_2) \widebar{\F}_{\aT_2}(q_2^2) + T_{\sT}^{\mu \nu \alpha}(-q_1,q_2) \widebar{\F}_{\sT}(q_2^2) \Big] \Big\rvert_{q_1^2 = \Mrho^2} \notag,
\end{align}
where we dropped an unobservable overall phase.

Spin-averaging the squared amplitude and performing the angular integration, we find 
\begin{align}
    \sigma(e^+ e^- \to f_1 \rho)(s)
    &= \frac{e^4 \lvert g_{\rho \gamma} \rvert^2 \lvert \vec{p}_\rho \rvert (s + 2m_e^2)}{384 \pi \Mf^6 \Mrho^4 s^3 \lvert \vec{p}_e \rvert} \notag \\
    &\times \Big[ T_{\aT_1,\aT_1}(q_1^2,s) \big\lvert \widebar{\F}_{\aT_1}(s) \big\rvert^2 + T_{\aT_2,\aT_2}(q_1^2,s) \big\lvert \widebar{\F}_{\aT_2}(s) \big\rvert^2 + T_{\sT,\sT}(q_1^2,s) \big\lvert \widebar{\F}_{\sT}(s) \big\rvert^2 \notag \\
    &\quad + 2 T_{\aT_1,\aT_2}(q_1^2,s) \Re\big[ \widebar{\F}_{\aT_1}(s) \widebar{\F}_{\aT_2}^*(s) \big] + 2 T_{\aT_1,\sT}(q_1^2,s) \Re\big[ \widebar{\F}_{\aT_1}(s) \widebar{\F}_{\sT}^*(s) \big] \notag \\
    &\quad + 2 T_{\aT_2,\sT}(q_1^2,s) \Re\big[ \widebar{\F}_{\aT_2}(s) \widebar{\F}_{\sT}^*(s) \big] \Big] \Big\rvert_{q_1^2 = \Mrho^2}
\end{align}
for the total cross section, with $s = q_2^2$, initial- and final-state momenta
\begin{align}
    \lvert \vec{p}_e \rvert
    &= \frac{\sqrt{s - 4m_e^2}}{2},
    &
    \lvert \vec{p}_\rho \rvert
    &= \frac{\sqrt{\lambda(s,q_1^2,\Mf^2)}}{2\sqrt{s}},
\end{align}
and the kinematic functions 
\begin{align}\label{eq:structures_cross_section_ee_f1_rho}
    T_{\aT_1,\aT_1}(q_1^2,s)
    &= 4 \big[ \lambda(s,q_1^2,\Mf^2) \big]^2, \\
    T_{\aT_2,\aT_2}(q_1^2,s)
    &= \Mf^6 (s + q_1^2) - \Mf^4 (s^2 + q_1^4 - 6 s q_1^2) - \Mf^2 (s - q_1^2)^2 (s + q_1^2) + (s - q_1^2)^4, \notag \\
    T_{\sT,\sT}(q_1^2,s)
    &= \Mf^2 (s + q_1^2) \big(\Mf^4-s^2 - q_1^4 + 18 s q_1^2\big) - \Mf^4 (s^2 + q_1^4 + 14 s q_1^2) + (s^2 - q_1^4)^2, \notag \\
    T_{\aT_1,\aT_2}(q_1^2,s)
    &= 2 \lambda(s,q_1^2,\Mf^2) \big[ (s - q_1^2)^2 - \Mf^2 (s + q_1^2) \big], \notag \\
    T_{\aT_1,\sT}(q_1^2,s)
    &= -2 \lambda(s,q_1^2,\Mf^2) (s - q_1^2) (s + q_1^2 - \Mf^2), \notag \\
    T_{\aT_2,\sT}(q_1^2,s)
    &= -(s - q_1^2) \Big[ \Mf^6 - \Mf^4 (s + q_1^2) - \Mf^2 (s^2 + q_1^4 - 6sq_1^2) + (s - q_1^2)^2 (s + q_1^2) \Big]. \notag
\end{align}
In a compact way, the cross section can be expressed in terms of the amputated helicity amplitudes~\cite{Hoferichter:2020lap}
\begin{align}\label{eq:helicity_amplitudes}
    \widebar{H}_{++;0}(q_1^2,s)
    &= \frac{\lambda(s,q_1^2,\Mf^2)}{2\Mf^3} \widebar{\F}_{1}(s) - \frac{q_1^2 (\Mf^2 - q_1^2 + s)}{2\Mf^3} \widebar{\F}_{2}(s) - \frac{s (\Mf^2 + q_1^2 - s)}{2\Mf^3} \widebar{\F}_{3}(s), \notag \\
    \widebar{H}_{+0;+}(q_1^2,s)
    &= \frac{q_1^2 s}{\xi_2 \Mf^2}\widebar{\F}_{2}(s) + \frac{s (\Mf^2 - q_1^2 - s)}{2\xi_2 \Mf^2} \widebar{\F}_{3}(s), \notag \\
    \widebar{H}_{0+;-}(q_1^2,s)
    &= -\frac{q_1^2 (\Mf^2 - q_1^2 - s)}{2\xi_1 \Mf^2} \widebar{\F}_{2}(s) - \frac{q_1^2 s}{\xi_1 \Mf^2} \widebar{\F}_{3}(s),
\end{align}
with polarization-vector normalizations $\xi_1^2 = q_1^2$, $\xi_2^2 = s$, leading to 
\beq\label{eq:cross_section_rho}
    \sigma(e^+ e^- \to f_1 \rho)(s)
    = \frac{e^4 \lvert g_{\rho \gamma} \rvert^2 \lvert \vec{p}_\rho \rvert (s + 2m_e^2)}{24 \pi \Mrho^4 s^3 \lvert \vec{p}_e \rvert} \sum_{\lambda} \big\lvert \widebar{H}_\lambda(\Mrho^2,s) \big\rvert^2,
\eeq
and the sum extends over the three amplitudes in \autoref{eq:helicity_amplitudes}. 

\subsection[$e^+ e^- \to f_1 \pi^+ \pi^-$]{$\boldsymbol{e^+ e^- \to f_1 \pi^+ \pi^-}$}
To obtain a reasonable threshold behavior, it is mandatory to go beyond the approximation of a narrow $\rho$ and instead consider the full amplitude $e^+ e^- \to f_1\pi^+ \pi^-$.
To this end, we use
\beq
    \M(\rho \to \pi^+ \pi^-)
    = g_{\rho \pi \pi} \eps_\mu(p_\rho) (p_- - p_+)^\mu
\eeq
to calculate the diagram shown in \autoref{fig:ee_f1_pipi}, leading to
\begin{align}
    \M(e^+ e^- \to f_1 \pi^+ \pi^-)
    &= \frac{e^2 g_{\rho \pi \pi}}{\widetilde{g}_{\rho \gamma}\Mf^2} \eps_\alpha^*(P) \frac{(p_- - p_+)_\mu}{q_1^2 - \Mrho^2 + \iu \Mrho \Gammarho} \frac{\overline{v}^s(p_1) \gamma_\nu u^r(p_2)}{q_2^2} \\
    &\hspace{-1.5cm} \times \Big[ T_{\aT_1}^{\mu \nu \alpha}(-q_1,q_2) \widebar{\F}_{\aT_1}(q_2^2) + T_{\aT_2}^{\mu \nu \alpha}(-q_1,q_2) \widebar{\F}_{\aT_2}(q_2^2) + T_{\sT}^{\mu \nu \alpha}(-q_1,q_2) \widebar{\F}_{\sT}(q_2^2) \Big] \notag,
\end{align}
where we again dropped an unobservable phase. 
From the spin-averaged squared matrix element and after carrying out the angular integrations, we obtain the differential cross section
\begin{align}\label{eq:dsig_f1_pipi}
    \frac{\mathrm{d}\sigma(e^+ e^- \to f_1 \pi^+ \pi^-)}{\mathrm{d}q_1^2}(s)
    &= \frac{e^4 \lvert g_{\rho \pi \pi} \rvert^2 \lvert g_{\rho \gamma} \rvert^2 \lvert \vec{p}_\rho \rvert  (s + 2m_e^2)(q_1^2 - 4\Mpi^2)^{3/2}}{18432\pi^3 \Mf^6 \Mrho^4 s^3 \sqrt{q_1^2} \lvert \vec{p}_e \rvert \big[ (q_1^2 - \Mrho^2)^2 + \Mrho^2 \Gammarho^2 \big]} \notag \\
    &\hspace{-0.5cm}\times \Big[ T_{\aT_1,\aT_1}(q_1^2,s) \big\lvert \widebar{\F}_{\aT_1}(s) \big\rvert^2 + T_{\aT_2,\aT_2}(q_1^2,s) \big\lvert \widebar{\F}_{\aT_2}(s) \big\rvert^2 + T_{\sT,\sT}(q_1^2,s) \big\lvert \widebar{\F}_{\sT}(s) \big\rvert^2 \notag \\
    &\hspace{-0.5cm}\quad + 2 T_{\aT_1,\aT_2}(q_1^2,s) \Re\big[ \widebar{\F}_{\aT_1}(s) \widebar{\F}_{\aT_2}^*(s) \big] + 2 T_{\aT_1,\sT}(q_1^2,s) \Re\big[ \widebar{\F}_{\aT_1}(s) \widebar{\F}_{\sT}^*(s) \big] \notag \\
    &\hspace{-0.5cm}\quad + 2 T_{\aT_2,\sT}(q_1^2,s) \Re\big[ \widebar{\F}_{\aT_2}(s) \widebar{\F}_{\sT}^*(s) \big] \Big],
\end{align}
with the kinematic functions $T_{i,j}(q_1^2,s)$ as in \autoref{eq:structures_cross_section_ee_f1_rho}.
In terms of the amputated helicity amplitudes, we obtain 
\beq\label{eq:cross_section_pipi}
    \frac{\mathrm{d}\sigma(e^+ e^- \to f_1 \pi^+ \pi^-)}{\mathrm{d}q_1^2}(s)
    = \frac{e^4 \lvert g_{\rho \pi \pi} \rvert^2 \lvert g_{\rho \gamma} \rvert^2 \lvert \vec{p}_\rho \rvert (s + 2m_e^2)(q_1^2 - 4\Mpi^2)^{3/2}}{1152 \pi^3 \Mrho^4 s^3 \sqrt{q_1^2} \lvert \vec{p}_e \rvert \big[ (q_1^2 - \Mrho^2)^2 + \Mrho^2 \Gammarho^2 \big]}
    \sum_{\lambda} \big\lvert \widebar{H}_\lambda(q_1^2,s) \big\rvert^2.
\eeq
In general, the remaining integration over $q_1^2$ needs to be performed numerically, but it is instructive to consider the limit of a narrow resonance~\cite{Hoferichter:2012pm}
\beq
    \frac{1}{(q_1^2 - \Mrho^2)^2 + \Mrho^2 \Gammarho^2}
    \to \frac{\pi}{\Mrho \Gammarho} \delta \big( q_1^2 - \Mrho^2 \big).
\eeq
In this approximation, together with
\beq\label{eq:rho_width}
    \Gammarho
    = \frac{\lvert g_{\rho \pi \pi} \rvert^2(\Mrho^2 - 4\Mpi^2)^{3/2}}{48 \pi \Mrho^2},
\eeq
the $q_1^2$ integration of \autoref{eq:cross_section_pipi} indeed reproduces \autoref{eq:cross_section_rho}.
For the phenomenological analysis of the $e^+ e^- \to f_1 \pi^+ \pi^-$ data, we will use the full expression given in \autoref{eq:dsig_f1_pipi}.

\section{Phenomenological analysis}
\label{sec:pheno}

\subsection[Data input for $e^+ e^- \to e^+ e^- f_1$ and $f_1 \to V \gamma$]{Data input for $\boldsymbol{e^+ e^- \to e^+ e^- f_1}$ and $\boldsymbol{f_1 \to V \gamma}$}
\label{sec:data_L3_Vgamma}
\begin{table}[t]
	\centering
	\begin{tabular}{l  r  r  r}
	\toprule
		Quantity & Value & Reference\\\midrule
		$\widetilde{\Gamma}_{\gamma\gamma}^{f_1} \ [\text{keV}]$ & $3.5(6)(5)$ & \cite{Achard:2001uu} \\
		$\Lambda_{f_1} \ [\text{GeV}]$ & $1.04(6)(5)$ & \cite{Achard:2001uu} \\
		$\BR(f_1 \to \rho \gamma)$ & $4.2(1.0)\perc$ & \cite{Zanke:2021wiq} \\
		$r_{\rho \gamma}$ & $3.9(1.3)$ & \cite{Amelin:1994ii} \\
		$\BR(f_1 \to \phi \gamma)$ & $0.74(26) \times 10^{-3}$ & \cite{Bityukov:1987bj,ParticleDataGroup:2022pth} \\
		\bottomrule
	\end{tabular}
	\caption{Data for $e^+ e^- \to e^+ e^- f_1$ and $f_1 \to V \gamma$ used in our analysis.}
	\label{tab:data_input}
\end{table}
The experimental data we will use for the space-like reaction $e^+ e^- \to e^+ e^- f_1$ and the radiative decays $f_1 \to V \gamma$ are summarized in~\autoref{tab:data_input}.
For the former, this concerns normalization and slope from the L3 experiment~\cite{Achard:2001uu}, with isoscalar corrections evaluated using the mixing angle that follows from a combined analysis with the analogous quantities for the $f_1(1420)$~\cite{Achard:2007hm}, see \autoref{sec:VMD}.
For $\BR(f_1 \to \rho \gamma)$, we use the results of the global fit from Ref.~\cite{Zanke:2021wiq}, including data on $\Gamma(f_1 \to K \bar K \pi)/\Gamma(f_1 \to 4\pi)$~\cite{WA76:1989wds,WA76:1989mye,WA102:1997gkz}, $\Gamma(f_1 \to 4\pi)/\Gamma(f_1 \to \eta \pi \pi)$~\cite{Amsterdam-CERN-Nijmegen-Oxford:1978mws,Bolton:1991nx}, $\Gamma(f_1 \to \rho \gamma)/\Gamma(f_1 \to 4\pi)$~\cite{MARK-III:1989jot}, $\Gamma(f_1 \to a_0(980) \pi \,[\text{excluding }\,K \bar K \pi])/\Gamma(f_1 \to \eta \pi \pi)$~\cite{Amsterdam-CERN-Nijmegen-Oxford:1978mws,Corden:1978cz,CLAS:2016zjy}, $\Gamma(f_1 \to K \bar K \pi)/\Gamma(f_1 \to \eta \pi \pi)$~\cite{Corden:1978cz,Amsterdam-CERN-Nijmegen-Oxford:1978mws,WA102:1998zhh,CLAS:2016zjy,Campbell:1969cx,Defoix:1972vd}, and $\Gamma(f_1 \to \rho \gamma)/\Gamma(f_1 \to \eta \pi \pi)$~\cite{CLAS:2016zjy,WA102:1998zhh,WA76:1991lql,Amelin:1994ii}.
As detailed in Ref.~\cite{Zanke:2021wiq}, our fit differs from the PDG average, $\BR(f_1 \to \rho \gamma) = 6.1(1.0)\perc$~\cite{ParticleDataGroup:2022pth}, for two main reasons: we include the result from Ref.~\cite{Amelin:1994ii}, which reduces the average, and we set the fit up in terms of $\Gamma(f_1 \to \rho \gamma)/\Gamma(f_1 \to \eta \pi \pi)$ instead of $\Gamma(f_1 \to \eta \pi \pi)/\Gamma(f_1 \to \rho \gamma)$ as in Ref.~\cite{ParticleDataGroup:2022pth}, since the latter introduces a bias towards larger branching fractions $\BR(f_1 \to \rho \gamma)$.
For $\BR(f_1 \to \phi \gamma)$, there is a single measurement from Ref.~\cite{Bityukov:1987bj}, and we will consider fit variants with and without this additional input, given both the tenuous data situation and the required $\Uthree$ assumptions.\footnote{%
    The limit $\BR(f_1 \to \phi \gamma) < 0.45 \times 10^{-3}$~\cite{WA102:1998zhh,ParticleDataGroup:2022pth} ($95\perc$ C.L.) supports a rather small branching fraction to $\phi \gamma$, indicating a value at the lower end of the range from Ref.~\cite{Bityukov:1987bj}.
    Both measurements are also consistent with $\BR(f_1 \to \phi \gamma) < 0.9 \times 10^{-3}$~\cite{Amelin:1994ii} ($95\perc$ C.L.).}
Finally, two event candidates for $f_1 \to e^+ e^-$ have been observed in Ref.~\cite{SND:2019rmq}, which, when interpreted as a signal, translates to $\BR(f_1 \to e^+ e^-) = 5.1^{+3.7}_{-2.7} \times 10^{-9}$, while being quoted as $\BR(f_1 \to e^+ e^-) < 9.4 \times 10^{-9}$ ($90\perc$ C.L.) in Ref.~\cite{ParticleDataGroup:2022pth}.
In Ref.~\cite{Zanke:2021wiq} we performed a detailed analysis of the constraints that can be obtained from the dilepton decay, but in view of its unclear status and large uncertainties, we no longer include this channel in our global fit here and instead focus on $e^+ e^- \to f_1 \pi^+ \pi^-$.
Further input parameters are collected in \autoref{appx:constants}.

\subsection[Data input for $e^+ e^- \to f_1 \pi^+ \pi^-$]{Data input for $\boldsymbol{e^+ e^- \to f_1 \pi^+ \pi^-}$}
\label{sec:data_ee_f1_pipi}
The process $e^+ e^- \to f_1 \pi^+ \pi^-$ has been measured in two different decay channels, $f_1 \to \eta \pi \pi$~\cite{BaBar:2007qju} and $f_1 \to K \bar K \pi $~\cite{BaBar:2022ahi}.
The data for the cross section from both reconstruction methods are well compatible, indicating that systematic errors are smaller than the statistical uncertainties of the measurements.
In the following, we will therefore assume that the data are indeed dominated by statistics.

Next, around $\sqrt{s} \simeq 2\GeV$, the cross section displays resonance structures~\cite{Liu:2022yrt}, most prominently the $\rho(2150)$, and, potentially, further excited $\rho$ states.
This implies that we cannot expect our theoretical description based on \autoref{eq:dsig_f1_pipi} to provide an adequate fit to the data, because $\rho$ excitations beyond the $\rho(1700)$ are not included.
However, the data still provide a valuable upper bound for the background contributions that our TFF parameterizations do describe; in fact, this constraint proves extremely stringent, immediately ruling out, by at least an order of magnitude, parameterizations that do not implement the doubly-virtual asymptotic behavior of \autoref{eq:VMD_asymptotics}.
Even more, writing the cross section in terms of the couplings $C_{\aT_{1/2}}$, $C_{\sT}$, one observes that moderate cancellations among the different terms are required to obey the upper limit implied by the $e^+ e^- \to f_1 \pi^+ \pi^-$ data.
With $C_{\sT}$ reasonably well determined from the L3 equivalent two-photon decay width, this thus implies a valuable constraint on the antisymmetric TFFs. 
\begin{table}[t]
	\centering
	{
   \setlength{\tabcolsep}{20pt}
	\begin{tabular}{l  r  r}
	\toprule
    & \multicolumn{2}{c}{$f_1 \to \phi \gamma$} \\
    & \multicolumn{1}{c}{No} & \multicolumn{1}{c}{Yes} \\
    \midrule	
    $\chi^2/\text{dof}$ & $5.6/3 = 1.86$ & $18.1/4 = 4.52$ \\
    $p$-value & $0.13$ & $1.2 \times 10^{-3}$ \\
    \midrule
    $C_\sT$ & $0.95(13)$ &$0.76(16)$ \\
    $C_{\aT_1}$ & $-0.16(18)$ & $-0.07(18)$ \\
    $C_{\aT_2}$ & $0.47(25)$ &$0.09(32)$ \\
    $\rho_{\sT \aT_1}$ & $0.34$ & $0.31$ \\
    $\rho_{\sT \aT_2}$ & $-0.11$ & $-0.34$ \\
    $\rho_{\aT_1 \aT_2}$ & $-0.52$ & $-0.35$ \\
    \midrule
    $\BR(f_1 \to \phi \gamma) \times 10^3$ & $3.4(1.7)$ & $1.6(1.0)$ \\
    $\BR(f_1 \to \omega \gamma) \times 10^3$ & $5.5(1.6)$ & $2.5(1.1)$ \\
    $\BR(f_1 \to e^+ e^-) \times 10^9$ & $2.2(6)$ & $1.2(5)$ \\
    $\BR(f_1' \to \phi \gamma) \times 10^3$ & $11.0(3.0)$ & $5.2(2.2)$ \\
    $\BR(f_1' \to \rho \gamma) \times 10^3$ & $4.8(2.6)$ & $2.2(1.4)$ \\
    \bottomrule
	\end{tabular}
	}
	\caption{Best-fit results for the three VMD couplings $C_\sT$, $C_{\aT_1}$, and $C_{\aT_2}$.
    The fit includes the constraints from the normalization and slope measured by L3 in $e^+ e^- \to e^+ e^- f_1$, from $\BR(f_1 \to \rho \gamma)$, $r_{\rho \gamma}$, and $\sigma(e^+ e^- \to f_1 \pi^+ \pi^-)$, as well as, in the right column, from $\BR(f_1 \to \phi \gamma)$.
    All uncertainties are inflated by the scale factor $S = \sqrt{\chi^2/\text{dof}}$.
    The table also shows the correlations $\rho_{ij}$ among the three couplings and the values of $\BR(f_1 \to V \gamma)$, $V = \omega,\phi$, and $\BR(f_1 \to e^+ e^-)$ implied by the fit result (the latter for $\sqrt{\sm} = 1.3\GeV$).
    The uncertainties for $\BR(f_1 \to V \gamma)$ include the fit errors and $\Delta R^\phi$, but no additional estimate of $\Uthree$ uncertainties.
    The predictions for $\BR(f_1' \to \phi \gamma)$ and $\BR(f_1' \to \rho \gamma)$ use the $\Uthree$ relations from \autoref{eq:R_f1prime}.}
	\label{tab:solutions_global_fit}
\end{table}

To quantify this constraint, we proceed as follows: we first define the $\chi^2$ function
\beq
    \chi^2_\text{BaBar}(C_{\sT},C_{\aT_{1}},C_{\aT_{2}})
    = \sum_{i=1}^{n_\text{BaBar}} \frac{(\sigma(s_i,C_{\sT},C_{\aT_{1}},C_{\aT_{2}}) - \sigma_i^\text{exp})^2}{(\Delta\sigma_i^\text{exp})^2} \theta\big[ \sigma(s_i,C_{\sT},C_{\aT_{1}},C_{\aT_{2}}) - \sigma_i^\text{exp} \big],
\eeq
where $n_\text{BaBar} = 52$ is the combined number of data points from Refs.~\cite{BaBar:2007qju,BaBar:2022ahi}, $\sigma_i^\text{exp}$ and $\Delta\sigma_i^\text{exp}$ are central value and error at center-of-mass energy $\sqrt{s_i}$, respectively, and the \LN{Heaviside} function demands that contributions to $\chi^2_\text{BaBar}$ only arise when the theoretical model exceeds the central value of the data, thus not penalizing a potential excess of the latter due to excited $\rho$ resonances.
Interpreting this $\chi^2$ function in the usual statistical sense, however, puts an undue emphasis on the $e^+ e^- \to f_1 \pi^+ \pi^-$ data, especially in view of the uncertainties from the contamination of resonant contributions.
For this reason, we instead study contours in the $C_{\aT_{1}}$--$C_{\aT_{2}}$ plane for which $\chi^2_\text{BaBar}/\text{dof} = 1$ at a given value of $C_{\sT}$, which should provide a reasonable measure of the consistency of the encompassed values of $C_{\aT_{1/2}}$ with the experimental constraints.
We repeat this procedure for the relevant range of $C_{\sT}$ and formulate the resulting constraint on $C_{\aT_{1/2}}$ in terms of an ellipse whose parameters are interpolated as a function of $C_{\sT}$.
The final constraint is then written as
\beq\label{eq:chi2_BaBar_eff}
    \chi^2_\text{BaBar, eff}(C_{\sT},C_{\aT_{1}},C_{\aT_{2}})
    = \Delta \boldsymbol{y}_{\aT}^\intercal C^{-1}_\aT \Delta \boldsymbol{y}_{\aT}, 
\eeq
where 
\beq
    \Delta \boldsymbol{y}_{\aT}
    =
    \begin{pmatrix}
        C_{\aT_{1}} - C_{\aT_{1}}^{(0)} \\ 
        C_{\aT_{2}} - C_{\aT_{2}}^{(0)}
    \end{pmatrix},
\eeq
with central values $C_{\aT_{1/2}}^{(0)}$ and covariance matrix $C_\aT$, determined via the $\chi^2_\text{BaBar}/\text{dof}=1$ contour ellipse, implicitly depending on $C_{\sT}$.
This effective $\chi^2$ function defined in~\autoref{eq:chi2_BaBar_eff} is then used as input in the global fit, counted as two data points in the number of degrees of freedom.
This procedure is further motivated by the fact that the constraints imposed by the cross-section measurements at different energies will be highly correlated, since, if the upper limit is fulfilled at some point $s_i$ for a set of couplings $C_{\aT_{1}}$, $C_{\aT_{2}}$, $C_{\sT}$, the smoothness of the cross section makes it likely that the same holds true at neighboring points as well.

\subsection{Global fit}
\label{sec:global_fit}
\begin{figure}[t]
	\centering
	\includegraphics[width=0.496\textwidth]{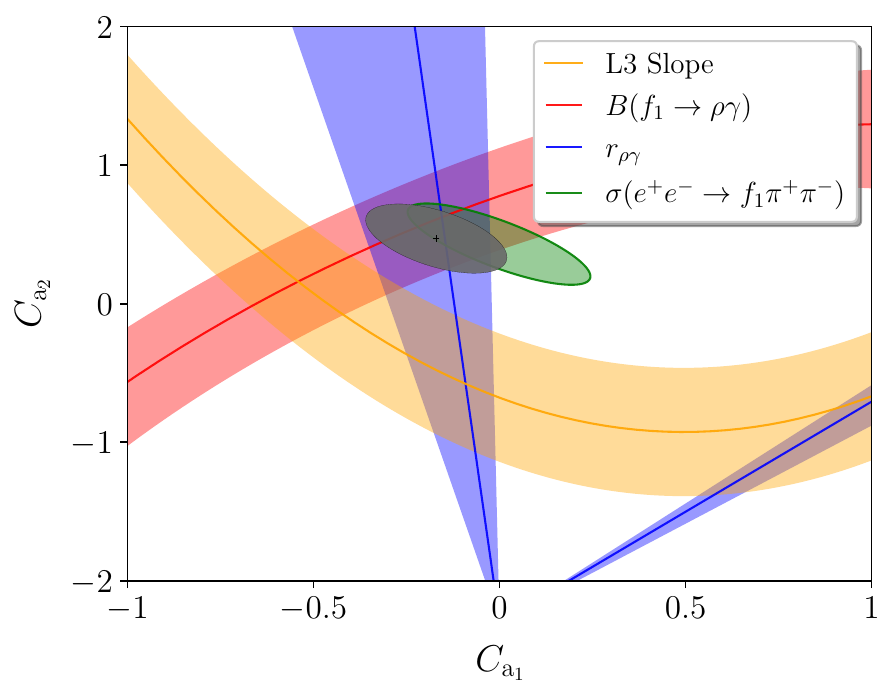}
	\includegraphics[width=0.496\textwidth]{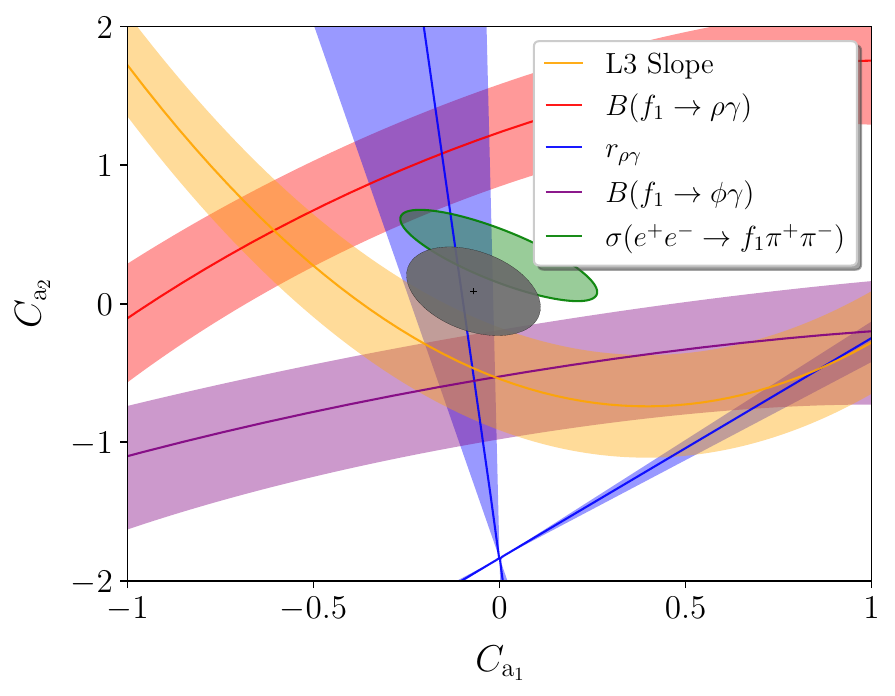}
	\caption{Left: constraints in the $C_{\aT_1}$--$C_{\aT_2}$ plane for the respective best-fit value of $C_{\sT}$ (see~\autoref{tab:solutions_global_fit}), from L3 normalization and slope, $\BR(f_1 \to \rho \gamma)$, $r_{\rho \gamma}$, and $\sigma(e^+ e^- \to f_1 \pi^+ \pi^-)$.
    The gray ellipse represents the result of the global fit.
    Right: same figure for the global fit including, in addition, $\BR(f_1 \to \phi \gamma)$.}
	\label{fig:solutions_global_fit_pipi}
\end{figure}
\begin{figure}[t]
	\centering
	\includegraphics[width=0.975\textwidth]{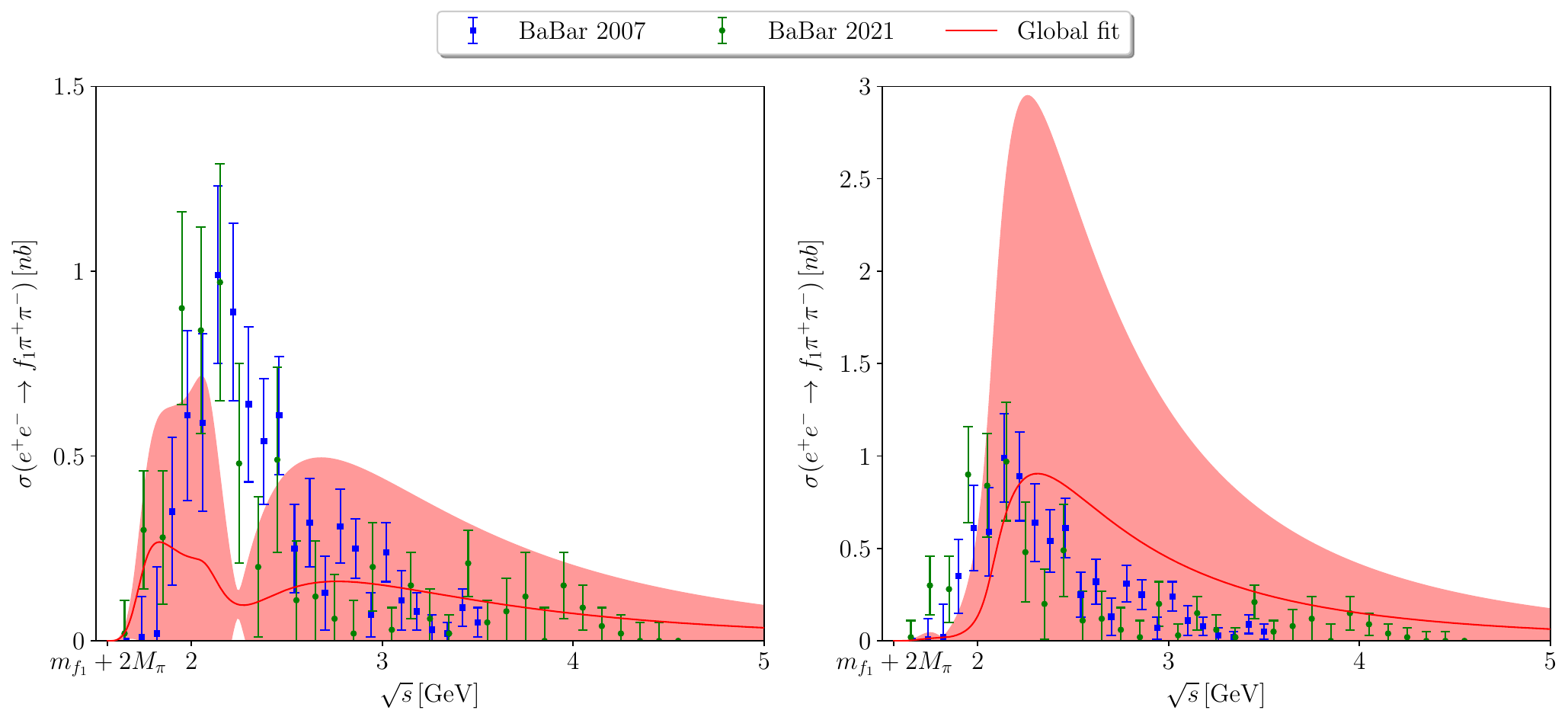}
	\caption{Comparison of our global fit results to the BaBar data~\cite{BaBar:2007qju,BaBar:2022ahi} for $\sigma(e^+ e^- \to f_1 \pi^+ \pi^-)$, without (left) and including (right) the constraint from $\BR(f_1 \to \phi \gamma)$.
    The red line denotes the central result and the band reflects the uncertainties propagated from $C_{\sT}$, $C_{\aT_1}$, and $C_{\aT_2}$.}
	\label{fig:f1_pipi_data_comparison}
\end{figure}

The results of the global fit are summarized in \autoref{tab:solutions_global_fit}, \autoref{fig:solutions_global_fit_pipi}, and \autoref{fig:f1_pipi_data_comparison}, for variants with and without the $\Uthree$ constraint from $\BR(f_1 \to \phi \gamma)$.
Without this input, we observe reasonable consistency among the various constraints, with a final value for $C_{\sT}$ close to the L3 value in~\autoref{eq:Cs_isoscalar}. 
The coupling $C_{\aT_1}$ comes out consistent with zero, while a non-zero value of $C_{\aT_2}$ is obtained at $2\sigma$ significance.
Crucially, owing to the inclusion of the BaBar data on $e^+ e^- \to f_1 \pi^+ \pi^-$~\cite{BaBar:2007qju,BaBar:2022ahi}, we are now able to provide an unambiguous solution for all three TFFs, including the two antisymmetric ones encoded in $C_{\aT_{1/2}}$.
The best-fit point lies within the ellipse from $\sigma(e^+ e^- \to f_1 \pi^+ \pi^-)$ and, accordingly, the central line in \autoref{fig:f1_pipi_data_comparison} respects the bound for almost all data points, leaving a deficit that could be well explained by a $\rho(2150)$-resonance signal.
Moreover, the resulting prediction for $\BR(f_1 \to e^+ e^-)$ is consistent with SND~\cite{SND:2019rmq}, suggesting a potential signal at the lower end of their range.
In contrast, the prediction for $\BR(f_1 \to \phi \gamma)$ comes out slightly too large in comparison to Ref.~\cite{Bityukov:1987bj}, in tension at the level of $1.5\sigma$. 

The same tension is visible in the global fit including $\BR(f_1 \to \phi \gamma)$, as the $\chi^2/\text{dof}$ deteriorates appreciably.
Including the resulting scale factor $S=2.1$ in the error estimates, all three couplings are consistent with the global fit without $\BR(f_1 \to \phi \gamma)$, but $C_{\sT}$ decreases compared to L3 and the central value of $C_{\aT_2}$ moves much closer to zero.
Within the sizable uncertainties, the cross section for $e^+ e^- \to f_1 \pi^+ \pi^- $ is still consistent, but the central line exceeds the data above the $\rho(2150)$, in accordance with the best-fit point in \autoref{fig:solutions_global_fit_pipi} lying slightly outside the $\sigma(e^+ e^- \to f_1 \pi^+ \pi^-)$ ellipse.
The resulting prediction for $\BR(f_1 \to e^+ e^-)$ is still consistent with SND, and $\BR(f_1 \to \phi \gamma)$ now agrees by construction.
\autoref{tab:solutions_global_fit} also includes the predictions for $\BR(f_1 \to \omega \gamma)$, $\BR(f_1' \to \phi \gamma)$, and $\BR(f_1' \to \rho \gamma)$, the latter two being related to the already determined couplings via the $\Uthree$ arguments in \autoref{sec:final_rep}. 

\subsection{Final representations}
\label{sec:final_rep}
To summarize, we propose that the low-energy contributions to the TFFs of the $f_1$ be described by the parameterizations
\begin{align}
    \F_{\aT_{1/2}}^{f_1, I=1}(q_1^2,q_2^2)
    &= \frac{R^\rho C_{\aT_{1/2}} \zeta_\rho \Mrho^4 \Mrhoprime^4 \Mrhoprimeprime^4 (q_1^2 - q_2^2)}{(q_1^2 - \Mrho^2)(q_2^2 - \Mrho^2)(q_1^2 - \Mrhoprime^2)(q_2^2 - \Mrhoprime^2)(q_1^2 - \Mrhoprimeprime^2)(q_2^2 - \Mrhoprimeprime^2)}, \notag \\
    \F_{\sT}^{f_1, I=1}(q_1^2,q_2^2)
    &= \frac{R^\rho C_{\sT} \Mrho^4 \Mrhoprime^4}{(q_1^2 - \Mrho^2)(q_2^2 - \Mrho^2)(q_1^2 - \Mrhoprime^2)(q_2^2 - \Mrhoprime^2)}, \notag \\
    \F_{\aT_{1/2}}^{f_1, I=0}(q_1^2,q_2^2)
    &= \sum_{V = \omega,\phi} \frac{R^V C_{\aT_{1/2}} \zeta_V \MV^4 \MVprime^4 \MVprimeprime^4 (q_1^2 - q_2^2)}{(q_1^2 - \MV^2)(q_2^2 - \MV^2)(q_1^2 - \MVprime^2)(q_2^2 - \MVprime^2)(q_1^2 - \MVprimeprime^2)(q_2^2 - \MVprimeprime^2)}, \notag \\
    \F_{\sT}^{f_1, I=0}(q_1^2,q_2^2)
    &= \sum_{V = \omega,\phi} \frac{R^V C_{\sT} \MV^4 \MVprime^4}{(q_1^2 - \MV^2)(q_2^2 - \MV^2)(q_1^2 - \MVprime^2)(q_2^2 - \MVprime^2)},
\end{align}
see \autoref{eq:VMD_isovector_inserted}, \autoref{eq:zetaV}, \autoref{eq:VMD_isoscalar}, and \autoref{eq:SU3_ratios_couplings}, with couplings $C_\sT$, $C_{\aT_1}$, and $C_{\aT_2}$ as determined in~\autoref{tab:solutions_global_fit} (and $R^\rho = 1$).
These low-energy contributions are then to be supplemented by the asymptotic contributions from the LCE, see \autoref{sec:TFFs}, to arrive at a complete description.

In order to estimate the impact of $f_1'$ and $a_1$, we also quote the corresponding expressions that follow from $\Uthree$ symmetry.
For the $f_1'$, the analogous results are obtained by replacing
\begin{align}\label{eq:R_f1prime}
    R^\rho
    &\to R_{f_1'}^\rho
    = \cot(\theta_\Ax + \theta_1)
    = -0.50(11), \notag \\
    R^\omega
    &\to R_{f_1'}^\omega
    = \frac{1}{9} \cot(\theta_\Ax + \theta_1)
    = -0.06(1), \notag \\
    R^\phi
    &\to R_{f_1'}^\phi
    = -\frac{2\sqrt{2}}{9}
    = -0.31,
\end{align}
where the errors only refer to the uncertainties propagated in $\theta_\Ax$, \Lat{cf.}\ \autoref{eq:thetaA} and \autoref{eq:SU3_ratios_couplings}.
The coefficients in \autoref{eq:R_f1prime} show that isoscalar contributions will become much more important for the $f_1'$ than for the $f_1$, especially the $\phi$.
This observation is reflected by some evidence for a signal in the decay to the $\phi \gamma$ final state, $\BR(f_1' \to \phi \gamma) = 3(2) \times 10^{-3}$~\cite{WA102:1998zhh}, which, within uncertainties, agrees with the predictions from \autoref{tab:solutions_global_fit} for the fit including $\BR(f_1\to\phi\gamma)$, while the fit without $\BR(f_1\to\phi\gamma)$ predicts a larger branching fraction.
The same reference also gives a limit $\BR(f_1' \to \rho \gamma) < 0.02$ ($95\perc$ C.L.), in agreement with both fits from \autoref{tab:solutions_global_fit}. 

The TFFs of the $a_1$ display a different isospin structure, with one isoscalar and one isovector photon each.
Moreover, for ideal mixing, there is no contribution from the $\phi$ and its excitations, so that only contributions of the $\rho$--$\omega$ type survive.
Accordingly, the overall scaling compared to the couplings in the $f_1$ TFFs is measured relative to the sum of all isovector and isoscalar contributions, leading to
\beq
    R_{a_1}
    = \frac{1 + R^\omega + R^\phi}{\sqrt{3} \cos(\theta_\Ax - \theta_0)}
    = \frac{2}{3 \sin(\theta_\Ax + \theta_1)}
    = 0.75(3).
\eeq
Choosing a symmetric decomposition of $\zeta_V$ onto the $\rho$ and $\omega$ contributions, we obtain 
\begin{align}
    \F_{\aT_{1/2}}^{a_1}(q_1^2,q_2^2)
    &= \frac{R_{a_1} C_{\aT_{1/2}} \sqrt{\zeta_\rho \zeta_\omega} \Mrho^2 \Mrhoprime^2 \Mrhoprimeprime^2 \Momega^2 \Momegaprime^2 \Momegaprimeprime^2 (q_1^2 - q_2^2)}{2(q_1^2 - \Mrho^2)(q_2^2 - \Momega^2)(q_1^2 - \Mrhoprime^2)(q_2^2 - \Momegaprime^2)(q_1^2 - \Mrhoprimeprime^2)(q_2^2 - \Momegaprimeprime^2)} \notag \\
    &+ (\rho \leftrightarrow \omega), \notag \\
    \F_{\sT}^{a_1}(q_1^2,q_2^2)
    &= \frac{R_{a_1} C_{\sT} \Mrho^2 \Mrhoprime^2 \Momega^2 \Momegaprime^2}{2(q_1^2 - \Mrho^2)(q_2^2 - \Momega^2)(q_1^2 - \Mrhoprime^2)(q_2^2 - \Momegaprime^2)} + (\rho \leftrightarrow \omega).
\end{align}

\section{Conclusions}
\label{sec:summary}
The transition form factors (TFFs) of axial-vector mesons are key input quantities for a data-driven evaluation of hadronic light-by-light (HLbL) scattering in the anomalous magnetic moment of the muon, yet they are notoriously poorly determined from experiment.
Here, we performed a global analysis of all experimental constraints available for the $f_1(1285)$ and outlined how the $f_1(1420)$ and $a_1(1260)$ contributions can be estimated from $\Uthree$ symmetry.
A crucial role is played by data for the cross section of $e^+ e^- \to f_1 \pi^+ \pi^-$, which provide valuable input on the asymptotic behavior and allowed us to find an unambiguous solution also for the antisymmetric TFFs. 

The process $e^+ e^- \to f_1 \pi^+ \pi^-$ probes all three TFFs at one photon virtuality determined by the center-of-mass energy and the other one by the $\pi^+ \pi^-$ invariant mass, which in turn is dominated by the $\rho(770)$.
Accordingly, the data extending from threshold up to about $4.5\GeV$ are sensitive to the asymptotic behavior for one virtuality fixed at the $\rho$ mass.
The corresponding constraint demonstrates that the asymptotic behavior predicted by the light-cone expansion needs to set in early, for otherwise, the cross section exceeds data by an order of magnitude.
We implemented this conclusion using a vector-meson-dominance ansatz, leading to the parameterizations summarized in \autoref{sec:final_rep}.
To account for contributions from even higher excited $\rho$ resonances, such as the $\rho(2150)$, we formulated the quantitative analysis as an upper limit, which still entails valuable constraints especially on the otherwise poorly determined couplings characterizing the antisymmetric TFFs.
The global fit, see \autoref{sec:global_fit}, shows good consistency with data for $e^+ e^- \to e^+ e^- f_1$ and $f_1 \to \rho \gamma$, predicting a branching fraction for $f_1 \to e^+ e^-$ at the lower end of the signal strength reported by SND.
Some tension is observed with $f_1 \to \phi \gamma$, which might point towards limitations of $\Uthree$ symmetry and/or the data base.

The final parameterizations describe the TFFs at low and intermediate virtualities, to be supplemented by an additional term from the light-cone expansion~\cite{Hoferichter:2020lap,Zanke:2021wiq} that ensures the correct asymptotic behavior also in the doubly-virtual direction.
Using this combined input, work is ongoing to evaluate the axial-vector contributions both in the HLbL basis of Ref.~\cite{Colangelo:2021nkr} and in the formalism of Ref.~\cite{Ludtke:2023hvz}.
In combination with the short-distance constraints from Refs.~\cite{Bijnens:2020xnl,Bijnens:2021jqo,Bijnens:2022itw}, the results presented here will thus be instrumental to arrive at a complete data-driven evaluation of HLbL scattering and to reduce the uncertainties to the level required by the final precision expected from the Fermilab experiment.


\acknowledgments
We thank Peter Stoffer for useful comments on the manuscript. 
Financial support by the DFG through the funds provided to the Sino--German Collaborative Research Center TRR110 ``Symmetries and the Emergence of Structure in QCD'' (DFG Project-ID 196253076 -- TRR 110) and the SNSF (Project No.\ PCEFP2\_181117) is gratefully acknowledged. 


\appendix

\section{Constants and parameters}
\label{appx:constants}
\begin{table}[!hpt]
	\centering
	\begin{tabular}{l  c  r  r}
	\toprule
		Quantity & Variable & Value & Reference\\
		\midrule
		Mass pion & $\Mpi$ & $139.57 \MeV$ & \multirow{19}{*}{\cite{ParticleDataGroup:2022pth}} \\
		Mass $f_1(1285)$ & $\Mf$ & $1281.9(5) \MeV$ & \\
		Mass $f_1(1420)$ & $\Mfprime$ & $1426.3(9) \MeV$ & \\
		Mass $a_1(1260)$ & $\Ma$ & $1230(40) \MeV$ & \\
		Mass $\omega(782)$ & $\Momega$ & $782.66(13) \MeV$ & \\
		Mass $\omega(1420)$ & $\Momegaprime$ & $1410(60) \MeV$ & \\
		Mass $\omega(1650)$ & $\Momegaprimeprime$ & $1670(30) \MeV$ & \\
		Mass $\phi(1020)$ & $\Mphi$ & $1019.461(16) \MeV$ & \\
		Mass $\phi(1680)$ & $\Mphiprime$ & $1680(20) \MeV$ & \\
		Mass $\phi(2170)$ & $\Mphiprimeprime$ & $2163(7) \MeV$ & \\
		Mass $\rho(770)$ (charged) & $\Mrho$ & $775.11(34) \MeV$ & \\
		Mass $\rho(1450)$ & $\Mrhoprime$ & $1465(25) \MeV$ & \\
		Mass $\rho(1700)$ & $\Mrhoprimeprime$ & $1720(20) \MeV$ & \\
		Total width $f_1(1285)$ & $\Gamma_{f_1}$ & $22.7(1.1) \MeV$ & \\
		Total width $f_1(1420)$ & $\Gamma_{f_1'}$ & $54.5(2.6)\MeV$ & \\
		Total width $a_1(1260)$ & $\Gamma_{a_1}$ & $(250 \ldots 600) \MeV$ & \\
		Total width $\rho(770)$ (charged) & $\Gammarho$ & $149.1(8) \MeV$ & \\
		Total width $\rho(1450)$ & $\Gammarhoprime$ & $400(60) \MeV$ & \\
		Total width $\rho(1700)$ & $\Gammarhoprimeprime$ & $250(100) \MeV$ & \\
		\midrule
		Mass $\rho(770)$ (charged) & $\Mrho$ & $774.9(6) \MeV$& \multirow{6}{*}{\cite{Belle:2008xpe}} \\
		Mass $\rho(1450)$ (charged) & $\Mrhoprime$ & $1428(30) \MeV$ & \\
		Mass $\rho(1700)$ (charged) & $\Mrhoprimeprime$ & $1694(98) \MeV$ & \\
		Total width $\rho(770)$ (charged) & $\Gammarho$ & $148.6(1.8) \MeV$ & \\
		Total width $\rho(1450)$ (charged) & $\Gammarhoprime$ & $413(58) \MeV$ & \\
		Total width $\rho(1700)$ (charged) & $\Gammarhoprimeprime$ & $135(62) \MeV$ & \\
        \midrule
        Mass $\rho(770)$ (neutral) & $\Mrho$ & $775.02(35) \MeV$&\multirow{6}{*}{\cite{BaBar:2012bdw}} \\
		Mass $\rho(1450)$ (neutral) & $\Mrhoprime$ & $1493(15) \MeV$ & \\
		Mass $\rho(1700)$ (neutral) & $\Mrhoprimeprime$ & $1861(17) \MeV$ & \\
		Total width $\rho(770)$ (neutral) & $\Gammarho$ & $149.59(67) \MeV$ & \\
		Total width $\rho(1450)$ (neutral) & $\Gammarhoprime$ & $427(31) \MeV$ & \\
		Total width $\rho(1700)$ (neutral) & $\Gammarhoprimeprime$ & $316(26) \MeV$ & \\
		\bottomrule
	\end{tabular}
	\caption{Masses and decay widths from Ref.~\cite{ParticleDataGroup:2022pth} as used in this work (first panel), in comparison to the $\rho(770)$, $\rho(1450)$, and $\rho(1700)$ parameters from Refs.~\cite{Belle:2008xpe,BaBar:2012bdw}.}
	\label{tab:constants}
\end{table}

In \autoref{tab:constants}, we collect the masses and decay widths used in this work, in large part taken from Ref.~\cite{ParticleDataGroup:2022pth}.
For most quantities, possible effects from isospin breaking can be safely neglected, but some ambiguity arises for the mass and width of the $\rho(770)$.
For the $e^+ e^- \to f_1 \pi^+ \pi^-$ process as the focus of this work, it would be natural to identify the $\rho$ parameters with the $\rho^0$, whose width is quoted at an appreciably lower value than for the charged channel.
However, here we follow the arguments from Ref.~\cite{Zanke:2021wiq}, observing that determinations sensitive also to the excited $\rho$ states both in the neutral~\cite{Belle:2008xpe} and charged mode~\cite{BaBar:2012bdw} tend to support the charged-channel values from Ref.~\cite{ParticleDataGroup:2022pth} and, therefore, use the latter ones throughout.
In particular, via \autoref{eq:rho_width}, this determines $\lvert g_{\rho \pi \pi} \rvert = 5.98$, in good agreement with dispersive determinations~\cite{Garcia-Martin:2011nna,Hoferichter:2017ftn}.
Similarly, the photon couplings are calculated from \autoref{eq:G_V_gamma} with the branching fractions from Ref.~\cite{ParticleDataGroup:2022pth}, leading to
\beq
    \lvert g_{\rho \gamma} \rvert
    = 4.96,
    \qquad 
    \lvert g_{\omega \gamma} \rvert
    = 16.51,
    \qquad
    \lvert g_{\phi \gamma} \rvert
    = 13.40.
\eeq
Finally, we quote the values for masses and decay widths of the axial-vector resonances from Ref.~\cite{ParticleDataGroup:2022pth}.
For the $a_1$, the (reaction-dependent) \LN{Breit}--\LN{Wigner} parameters can also be compared to attempts to extract the pole position from $\tau \to 3\pi \nu_\tau$ data, $\sqrt{s_{a_1}} = \Ma - \iu\Gamma_{a_1}/2 = \big[1209(4)( ^{+12}_{-9}) - \iu 288(6)(^{+45}_{-10})\big]\MeV$~\cite{JPAC:2018zwp}.
In addition, Ref.~\cite{ParticleDataGroup:2022pth} quotes the average $\Gamma_{a_1} = 420(35)\MeV$~\cite{dArgent:2017gzv,COMPASS:2018uzl}, in line with the center of the estimated range quoted in \autoref{tab:constants}.
Based on the same two references, one would conclude the mass average $\Ma = 1250(20)\MeV$. 


\bibliographystyle{apsrev4-1_mod_2}
\bibliography{ref}

\end{document}